\documentclass[conference]{IEEEtran}

\usepackage{textcomp}
\usepackage{microtype}
\usepackage{csquotes}

\usepackage[table,dvipsnames]{xcolor} 
\definecolor{charcoal}{RGB}{54, 69, 79}
\definecolor{lightgray}{RGB}{242, 242, 242}
\definecolor{medgray}{RGB}{220, 220, 220}
\definecolor{paleblue}{HTML}{F7F7FF}

\usepackage{amsmath,amssymb,amsfonts}
\usepackage{algorithmic}

\usepackage{graphicx}
\usepackage{booktabs}
\graphicspath{ {./figures/} }

\usepackage{tikz}
\usetikzlibrary{shapes.geometric, arrows.meta, positioning, fit, backgrounds, calc}
\usepackage{pgfplots}
\usepackage{pgfplotstable}
\pgfplotsset{compat=1.18}

\usepackage{cite}
\usepackage{enumitem}
\usepackage{fontawesome5}
\usepackage[skins]{tcolorbox}
\usepackage{soul}

\usepackage{url}
\usepackage[hidelinks]{hyperref}

\title{Sustainability Is Not Linear: Quantifying Performance, Energy, and Privacy Trade-offs in On-Device Intelligence}

\author{
\IEEEauthorblockN{
Eziyo Ehsani\IEEEauthorrefmark{1},
Luca Giamattei\IEEEauthorrefmark{1},
Ivano Malavolta\IEEEauthorrefmark{2},
Roberto Pietrantuono\IEEEauthorrefmark{1}
}
\IEEEauthorblockA{\IEEEauthorrefmark{1}University of Naples Federico II, Naples, Italy}
\IEEEauthorblockA{\IEEEauthorrefmark{2}Vrije Universiteit Amsterdam, Amsterdam, The Netherlands}
}

\begin{document}

\maketitle

\begin{abstract}
The migration of Large Language Models (LLMs) from cloud clusters to edge devices promises enhanced privacy and offline accessibility, but this transition encounters a harsh reality: the physical constraints of mobile batteries, thermal limits, and, most importantly, memory constraints. To navigate this landscape, we constructed a replicable and reproducible experimental pipeline to profile the complex interplay between energy consumption, latency, and quality of LLMs on mobile devices. 

We harness this pipeline to conduct an empirical case study on a flagship Android device, 
capturing granular 
metrics across eight LLMs ranging from 0.5B to 9B parameters without requiring root access, ensuring our findings reflect realistic user conditions. 
The findings 
highlight the trade-offs between generation quality, performance, power and resource consumption, revealing which LLMs offer the best balance across metrics and under different conditions. Besides, we uncovered a counter-intuitive quantization-energy paradox: while modern importance-aware quantization successfully reduces memory footprints to fit larger models into RAM, we found it yields negligible energy savings compared to standard mixed-precision methods. This proves that for battery life, the architecture of the model, not its quantization scheme, is the decisive factor. We further identified that Mixture-of-Experts (MoE) architectures defy the standard size-energy trend, offering the storage capacity of a 7B model while maintaining the lower energy profile of a 1B to 2B model. Finally, an analysis of these multi-objective trade-offs reveals a pragmatic sweet spot of mid-sized models, such as Qwen2.5-3B, that effectively balance response quality with sustainable energy consumption.
\end{abstract}

\section{Introduction}
\label{sec:introduction}

Large Language Models (LLMs) are increasingly being integrated into interactive systems that must operate under latency, privacy, and availability constraints. This trend has increased interest in edge AI, where inference is executed directly on end-user devices such as smartphones, wearables, and embedded systems. For many applications, local execution is not only desirable but necessary. Edge inference becomes relevant when network connectivity is unreliable or unavailable, when user data is privacy-sensitive and cannot be transmitted to external cloud servers, or when responses must be delivered with low latency to support interactive use. Furthermore, offloading computation to the edge can reduce cloud infrastructure costs and server-side energy consumption by distributing computation across decentralized consumer devices~\cite{shi2016edge}. This decentralization may also reduce the carbon footprint associated with centralized data centers dedicated to generative inference~\cite{wu2022sustainable}.

However, moving from cloud infrastructure to mobile hardware fundamentally changes the feasibility constraints of deploying these models. Smartphones operate under limited compute throughput, memory bandwidth, thermal headroom, and finite battery capacity. LLM inference is challenging in this setting because it combines large memory footprints, required to store multi-billion-parameter model weights and the growing KV cache, with sustained compute demand during autoregressive decoding. This token generation phase is heavily \textbf{memory-bound}, meaning that the speed of data transfer between RAM and the processor often limits performance more than raw computational capacity~\cite{pope2023efficiently}. The gap between processor speed and memory bandwidth makes transferring multi-billion-parameter matrices for each generated token a major hardware bottleneck~\cite{aminabadi2022deepspeed}. As a result, a model that is accurate but drains the battery or triggers thermal throttling can become impractical for deployment, regardless of its generative quality or theoretical capabilities.

Understanding these trade-offs requires reliable energy profiling, yet accurate power measurement on consumer devices remains difficult. Traditional approaches rely on external power monitors, such as the Monsoon hardware power monitor~\cite{monsoon_hvpm}, which provide high-fidelity electrical measurements directly from the power source. However, these techniques typically require invasive hardware modifications, such as bypassing the internal battery, and are therefore not well aligned with evaluating intact consumer phones under realistic usage conditions. Conversely, software-only estimators based on CPU frequency states are easier to deploy but can be unreliable for long-running LLM workloads. Processor frequency is influenced by thermal throttling, operating system power-management policies, and background system activity, meaning that static power models can lose accuracy as the device heats up~\cite{mccullough2011evaluating}. As modern mobile operating systems continuously adjust voltage and frequency to prevent hardware damage, relying only on software heuristics can lead to estimation errors during sustained inference sessions~\cite{yoon2012appscope}. Consequently, many studies evaluate latency or compression techniques in isolation, leaving the \textbf{joint relationship between energy consumption, memory feasibility, performance, and generation quality} under-explored on mobile hardware.

In this paper, we address this gap by proposing a reproducible, non-intrusive measurement pipeline that estimates workload-attributable energy directly from native system APIs, while preserving realistic device behavior. We then use this pipeline to conduct an empirical case study evaluating on-device LLM inference on an unrooted flagship Android device, the Samsung Galaxy S25 Ultra. Using \texttt{llama.cpp} as the execution runtime~\cite{llamacpp}, we benchmark eight open-source LLMs spanning 0.5B to 9B parameters. The selection includes both dense architectures and sparse Mixture-of-Experts (MoE) architectures. We evaluate these models under two 4-bit quantization schemes: the mixed-precision \texttt{Q4\_K\_M} format and the importance-aware \texttt{IQ4\_XS} format. We evaluate each hardware-software configuration across multiple dimensions, including \textit{prefill speed, generation speed, overall latency}, and \textit{Time to First Token} (TTFT). We also measure resource usage through \textit{peak memory} footprints, evaluate energy efficiency via \textit{total Joules} and \textit{Joules per token}, and assess output quality using both reference-based evaluation with \textit{BERTScore}~\cite{bert_score} and reference-free evaluation with \textit{G-Eval}~\cite{G-Eval}. The case study addressed the following questions:

\begin{tcolorbox}[
    colback=paleblue,
    colframe=black,
    boxrule=0.5pt,
    left=1pt, right=1pt, top=1pt, bottom=1pt,
    ]
\faSearch\ \textbf{RQ$_1$.} 
 Is on-device LLM inference practically feasible on a flagship smartphone?
\end{tcolorbox}
This RQ examines the feasibility of on-device LLM inference in a high-end smartphone setting, considering RAM capacity, thermal limits, battery capacity, and latency. The objective is to determine whether, on the evaluated device, local execution of models ranging from 0.5B to 9B parameters can move beyond technical feasibility and achieve practical usability \textbf{without hardware failure, severe thermal throttling, or excessive user friction}.

\begin{tcolorbox}[
    colback=paleblue,
    colframe=black,
    boxrule=0.5pt,
    left=1pt, right=1pt, top=1pt, bottom=1pt,
    ]
\faSearch\ \textbf{RQ$_2$.} To what extent does quantization matter for on-device LLMs?
\end{tcolorbox}

RQ$_2$ examines the effect of different compression algorithms on mobile hardware. We investigate whether importance-aware formats translate into energy savings and latency reductions. By contrasting standard mixed-precision methods with codebook-based quantization, the analysis evaluates whether runtime weight reconstruction introduces processing overhead that \textbf{offsets theoretical efficiency benefits}.

\begin{tcolorbox}[
    colback=paleblue,
    colframe=black,
    boxrule=0.5pt,
    left=1pt, right=1pt, top=1pt, bottom=1pt,
    ]
\faSearch\ \textbf{RQ$_3$.} How do different model sizes and foundational architectures affect performance on mobile hardware?
\end{tcolorbox}
RQ$_3$ examines how different LLMs perform and how model size and architecture affect mobile inference during decoding. The analysis evaluates how parameter count increases memory traffic and whether sparse architectures, specifically Mixture-of-Experts, can \textbf{decouple active compute from total parameter capacity}, thereby changing the expected relationship between model size, performance, and energy cost on resource-constrained devices.

The paper makes three primary contributions: 
\begin{tcolorbox}[
    colback=black!5,
    colframe=white,
    boxrule=0.5pt,
    left=1pt, right=1pt, top=1pt, bottom=1pt,
    title={\faList\ \textbf{Contributions to Sustainable Edge Intelligence}},
    colbacktitle=black!80,
    coltitle=white,
    fonttitle=\bfseries,
    boxed title style={colframe=black!80}
]
\begin{itemize}[noitemsep, topsep=0pt, parsep=0pt, partopsep=0pt, leftmargin=*]
\item We provide a methodology for \textbf{reproducible and replicable energy, performance, resource usage, and quality profiling} of LLMs on unrooted Android systems, delivering an automated pipeline requiring no rooting or hardware modification.\footnote{For the sake of Open Science, we provide a replication package including the experimental pipeline, raw measurement data, analysis scripts, and complete results, available at: {\url{https://github.com/eziyoo/LLMs-on-Devices}}.}

\item We conduct a \textbf{joint quality and efficiency analysis} through an empirical \textbf{case study} on a flagship Android device. This analysis investigates the trade-offs among energy, speed, resource consumption, and generation quality. It also examines how reference-based metrics can misrepresent extractive generation behavior and shows how reference-free evaluation protocols complement them by better capturing summary quality.


\item We provide a \textbf{quantization and architecture trade-off characterization} across model sizes and architectures, contrasting dense models with Mixture-of-Experts and directly comparing the \texttt{Q4\_K\_M} and \texttt{IQ4\_XS} quantization paradigms in terms of latency, energy, and peak memory demands.

\end{itemize}
\end{tcolorbox}

Overall, these empirical evaluations provide practical implications for sustainable edge deployment. By analyzing the interplay between compression, architecture, and hardware constraints, we provide software engineers and researchers with guidelines for balancing model scale and quantization strategies against battery and thermal constraints on consumer mobile devices.

\section{Related Work}
\label{sec:related_work}
This section reviews work related to sustainable on-device LLM inference. We organize the discussion around five themes: energy efficiency in NLP and deployment strategies, model quantization, on-device inference systems, mobile energy measurement, and automated experimentation. 

\subsection{\textbf{Energy Efficiency in NLP and Deployment Strategies}}
Empirical work on sustainable NLP has examined both model-side efficiency techniques and system-side deployment choices. The Green AI movement, introduced by Schwartz et al.~\cite{schwartz2020green}, advocated for reporting efficiency metrics alongside accuracy during training and inference. Strubell et al.~\cite{strubell2019energy} highlighted the carbon footprint associated with training and deploying dense Transformer models, motivating research on efficient inference beyond data centers. The computational cost of Transformer architectures makes energy efficiency a key requirement for deployment outside cloud environments. Yuan et al.~\cite{yuan2024impact} studied Knowledge Distillation as a way to reduce inference cost at the architectural level. By comparing large Transformer models, such as BERT and GPT-2 style architectures, against distilled variants, they showed that transferring representations from a large teacher to a compact student changes the execution profile. They reported that distillation can reduce energy consumption while improving inference time, showing that it is an effective method for lowering compute demand without specialized hardware accelerators.

Other studies focus on where inference executes. The choice between cloud-based and edge-based execution involves trade-offs between network transmission costs and local computation costs. Nguyen et al.~\cite{nguyen2025ondevice} compared client-side energy usage when fetching LLM-generated content from a remote server versus generating equivalent text on-device. Across multiple LMs and content lengths, they found remote fetching to be more energy-efficient for the client device in most scenarios. They attributed this energy gap mainly to the high compute cost of local autoregressive generation and the correlation between execution time and total energy consumption on mobile systems. This highlights a tension in edge AI: local execution supports privacy and offline availability, but it increases the local power burden.

\subsection{\textbf{Model Quantization for LLMs}}
Running large language models locally is attractive for privacy and data sovereignty, but edge deployment is limited by memory, computation, and energy constraints. Abstreiter et al.~\cite{Maximilian} evaluated the feasibility of local generative inference on single-board computers using a short-story summarization task. They found that while quantization reduces memory overhead, resource bottlenecks still affect the on-device user experience. Sustained local generation led to noticeable latency, hardware overheating, and high energy consumption, highlighting the need for efficiency techniques that improve usability rather than only demonstrating technical feasibility.

Post-training quantization has become a main enabler of on-device LLM deployment. By reducing model weights from 16-bit floating-point representations to 8-bit or 4-bit integer formats, quantization mitigates the memory-bandwidth bottleneck that dominates generation. Approaches such as SmoothQuant~\cite{smoothquant} and GPTQ by Frantar et al.~\cite{frantar2022gptq} allow multi-billion-parameter models to fit within the memory budgets of mobile SoCs. Dettmers et al.~\cite{dettmers2022llmint8} showed that managing activation outliers is important for preventing quality degradation at low precision, further complicating deployment.

More recently, importance-aware methods such as Activation-aware Weight Quantization~\cite{AWQ} protect a small subset of sensitive weights using activation statistics from calibration datasets to reduce quantization error at low bit widths. However, most quantization studies emphasize memory footprint reductions and zero-shot accuracy metrics. The physical effects of different quantization layouts on mobile CPUs, especially sustained energy consumption and decoding latency, are less well characterized. This is particularly relevant for codebook and indirection-based formats such as \texttt{IQ4\_XS}, which require additional memory lookups and runtime reconstruction during inference.

\subsection{\textbf{On-Device and Edge LLM Inference Systems}}
The migration of LLM inference from cloud to edge devices has accelerated due to privacy concerns and the need for low-latency interaction. Early edge inference relied mainly on small models or aggressive compression that reduced model capability. More recent runtimes, such as \texttt{llama.cpp}~\cite{llamacpp}, and cross-platform compilation frameworks, such as Apache TVM~\cite{chen2018tvm}, have shown that quantized multi-billion-parameter models can run on consumer devices.

Benchmarking studies have started to characterize this setting. Laskaridis et al.~\cite{Mobile_transformers} measured throughput and latency across mobile CPUs and NPUs, emphasizing hardware bottlenecks such as limited memory bandwidth and the sequential nature of autoregressive decoding. Algorithmic improvements such as FlashAttention by Dao et al.~\cite{dao2022flashattention} show that input-output-aware inference can reduce memory reads and writes, although integrating these optimizations into quantized mobile CPU kernels remains a systems challenge. While prior work establishes technical feasibility and characterizes generation speed, physical energy consumption is often treated as secondary or approximated through software estimators. In addition, sparse architectures such as Mixture-of-Experts introduce a different compute-memory trade-off: they reduce active compute by routing tokens to selected subnetworks, but require all experts to remain resident in memory. This architectural paradigm remains under-explored under mobile execution constraints.

\subsection{\textbf{Mobile Energy Measurement}}
Accurate and reproducible energy measurement on commercial smartphones is difficult because of restricted telemetry access, thermal dynamics, and background operating system activity. Foundational studies such as Carroll and Heiser~\cite{carroll2010analysis} showed that component-level power draw is non-linear and context-dependent. External power monitors, such as the Monsoon hardware power monitor~\cite{monsoon_hvpm}, are often treated as the measurement reference because they capture electrical power directly at the source with high temporal resolution. However, they typically require disassembling the device, bypassing the internal battery, and using invasive hardware setups. This makes them difficult to apply in consumer-device studies and can alter the thermal behavior of the phone, making the setting less representative of real-world conditions.

Software-only approaches estimate energy indirectly by tracking CPU frequency states and using pre-computed device power models. While scalable, these estimators can be inaccurate for long-running and variable workloads such as LLM inference. Pathak et al.~\cite{pathak2011where} highlighted the difficulty of software-based energy accounting, noting that hidden system-call overheads and hardware tail-states can confound power models. During extended generation tasks, thermal throttling alters the relationship between frequency and power, while leakage current increases with processor temperature, causing static software estimators to diverge from physical measurements. Recent research has therefore explored native on-device APIs, such as Android \texttt{BatteryManager}~\cite{battery-manager}, to log voltage and current non-intrusively. This approach preserves realistic device behavior, but requires careful experimental controls, including background process isolation, consistent screen state management, 
and repeated runs.

\subsection{\textbf{Automated Experimentation and Measurement}}
Ensuring reproducibility is a long-standing challenge in empirical mobile systems and software engineering research. Green software engineering, pioneered by Hindle et al.~\cite{hindle2012green}, established the need for automated profiling to reveal the energy behavior of software execution. Mobile operating systems are dynamic, and small environmental variations can affect performance profiling. Orchestration frameworks such as Android Runner~\cite{malavolta2020android} and Experiment Runner~\cite{experiment-runner} address this by automating experiment execution and standardizing data collection. Android Runner targets Android applications and supports declarative experiment definitions, enabling consistent setup and teardown. Experiment Runner generalizes this approach into a platform-agnostic Python framework emphasizing automation, artifact persistence, and crash recovery.

Both orchestration systems adopt plugin architectures to integrate profilers and measurement tools. For example, Android Runner supports mobile-oriented profilers such as Monsoon~\cite{monsoon_hvpm} and Trepn~\cite{qualcomm_trepn}, while Experiment Runner is integrated with software-based carbon and energy estimators such as CodeCarbon~\cite{schmidt2021codecarbon} and EnergiBridge~\cite{sallou2023energibridge}. These frameworks motivate the controls used in our pipeline, including automated initialization, repeated execution, workload isolation, thermal cool-down periods, and structured artifact collection.

\emph{Positioning of this work.}
Building on these lines of research, our study provides a joint  evaluation of energy consumption, performance, resource usage, and output quality for on-device LLM inference. We measure prefill and generation speed, time to first token, end-to-end latency, peak memory, energy consumption, and linguistic output quality across multiple model scales and architectures. 
We also isolate the effects of compression by comparing a standard mixed-precision 4-bit quantization scheme with an importance-aware 4-bit scheme, showing the practical limits of algorithmic efficiency optimizations on mobile hardware.

\section{Methodology}
\label{sec:methodology}

This section describes the experimental design used to profile on-device large language model inference in an Android environment. We detail the selected models and quantization schemes, the hardware and software configurations, the execution pipeline and reproducibility controls, the energy measurement and post-processing procedures, and the evaluation metrics used to quantify system performance and generation quality.

\subsection{\textbf{Models and Quantization}}
\label{subsec:models_quant}

To capture realistic mobile trade-offs across the current open-source ecosystem, we evaluated a diverse set of instruction-tuned large language models spanning a parameter scale from 0.5B to 9B. This range was selected because models smaller than half a billion parameters often lack the reasoning capacity required for meaningful human assistance, whereas models exceeding 9B parameters typically surpass the unified memory limits of contemporary mobile devices, leading to out-of-memory errors or reliance on slow storage swap space. Our selection includes both standard dense Transformer models and a sparse Mixture-of-Experts model.

Candidate models were selected based on their rankings in the Hugging Face Open LLM Leaderboard~\cite{LeaderBoard}. To obtain coverage across varying computational scales, we selected the highest-ranked model within each approximate one billion parameter band up to the 9B limit. We prioritized \textbf{instruction-tuned base variants} rather than task-specific fine-tunes. Instruction-tuned models are trained to follow human directions and engage in dialogue~\cite{ouyang2022training,wei2021finetuned}, making them representative of general-purpose on-device digital assistants. Models falling within the 4B to 6B parameter range were excluded because the publicly available candidates in this band were lower-ranked in benchmark quality than several smaller models, making them less informative for a systematic, size-based empirical comparison. Table~\ref{tab:models} lists all evaluated models along with their architectural specifications.

\begin{table}[t]
\centering
\caption{Evaluated models and architectures.}
\label{tab:models}
\small
\renewcommand{\arraystretch}{1.3}
\begin{tabular}{lclcc}
\hline
\rowcolor{charcoal}
\textcolor{white}{\textbf{Model}} & \textcolor{white}{\textbf{Params (B)}} & \textcolor{white}{\textbf{Arch}} & \textcolor{white}{\textbf{Format}} & \textcolor{white}{\textbf{Ref}} \\
\hline
Qwen2-0\_5B       & 0.494 & Dense  & GGUF & \cite{Qwen2}   \\
\rowcolor{lightgray!40}
Qwen2.5-1.5B      & 1.5   & Dense  & GGUF & \cite{qwen2.5} \\
Phi-2             & 2.78  & Dense  & GGUF & \cite{phi2}    \\
\rowcolor{lightgray!40}
Qwen2.5-3B        & 3.086 & Dense  & GGUF & \cite{qwen2.5} \\
OLMoE-1B-7B-0125  & 6.919 & MoE & GGUF & \cite{olmoe}   \\
\rowcolor{lightgray!40}
Qwen2.5-7B        & 7.616 & Dense  & GGUF & \cite{qwen2.5} \\
Meta-Llama-3.1-8B & 8.03  & Dense  & GGUF & \cite{llama3}  \\
\rowcolor{lightgray!40}
Gemma-2-9B        & 9.0   & Dense  & GGUF & \cite{gemma2}  \\
\hline
\end{tabular}
\end{table}

All selected models were executed using the \texttt{llama.cpp}~\cite{llamacpp} inference engine, an optimized framework designed for resource-constrained edge environments. A standard 7B parameter model stored in 16-bit floating-point precision requires over 14 gigabytes of memory, which exceeds the RAM available to user-space applications on most flagship smartphones~\cite{ignatov2018ai}. Therefore, we quantized each model using the \texttt{llama-quantize}~\cite{llama_cpp_quantize} utility into two widely used 4-bit formats supported by the GGUF ecosystem.

The first format evaluated is \texttt{Q4\_K\_M}, a block-wise mixed-precision post-training quantization approach. Prior research shows that uniform quantization below 4 bits in block-wise methods can lead to substantial degradation in model quality and increased perplexity, whereas exceeding 6 bits increases memory usage beyond feasible smartphone RAM capacities~\cite{Mobile_transformers,dettmers2022optimizers}. The \texttt{Q4\_K\_M} scheme provides a practical architectural balance by preserving the most sensitive parts of the neural network, especially attention layers and feed-forward network components, at higher 6-bit precision. At the same time, it compresses less sensitive and redundant weights to 4-bit precision. This hybrid strategy maintains overall model quality and semantic reasoning capabilities while keeping the memory footprint low enough for on-device deployment.

\begin{figure}[htbp]
    \centering
    \includegraphics[width=\linewidth]{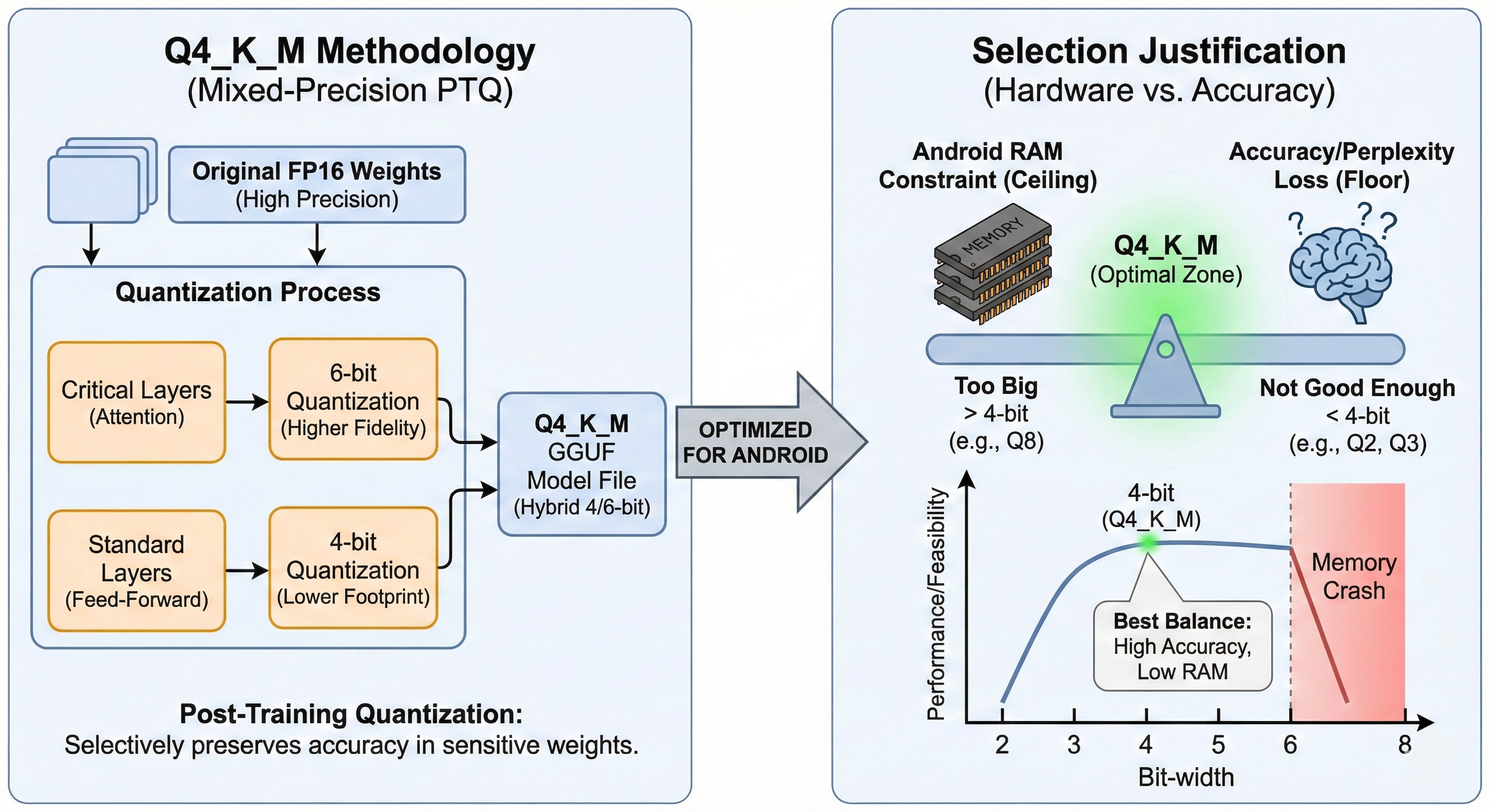}
    \caption{\texttt{Q4\_K\_M}: block-wise mixed precision with higher precision for more sensitive components.}
    \label{fig:q4km}
\end{figure}

\begin{figure}[htbp]
    \centering
    \includegraphics[width=\linewidth]{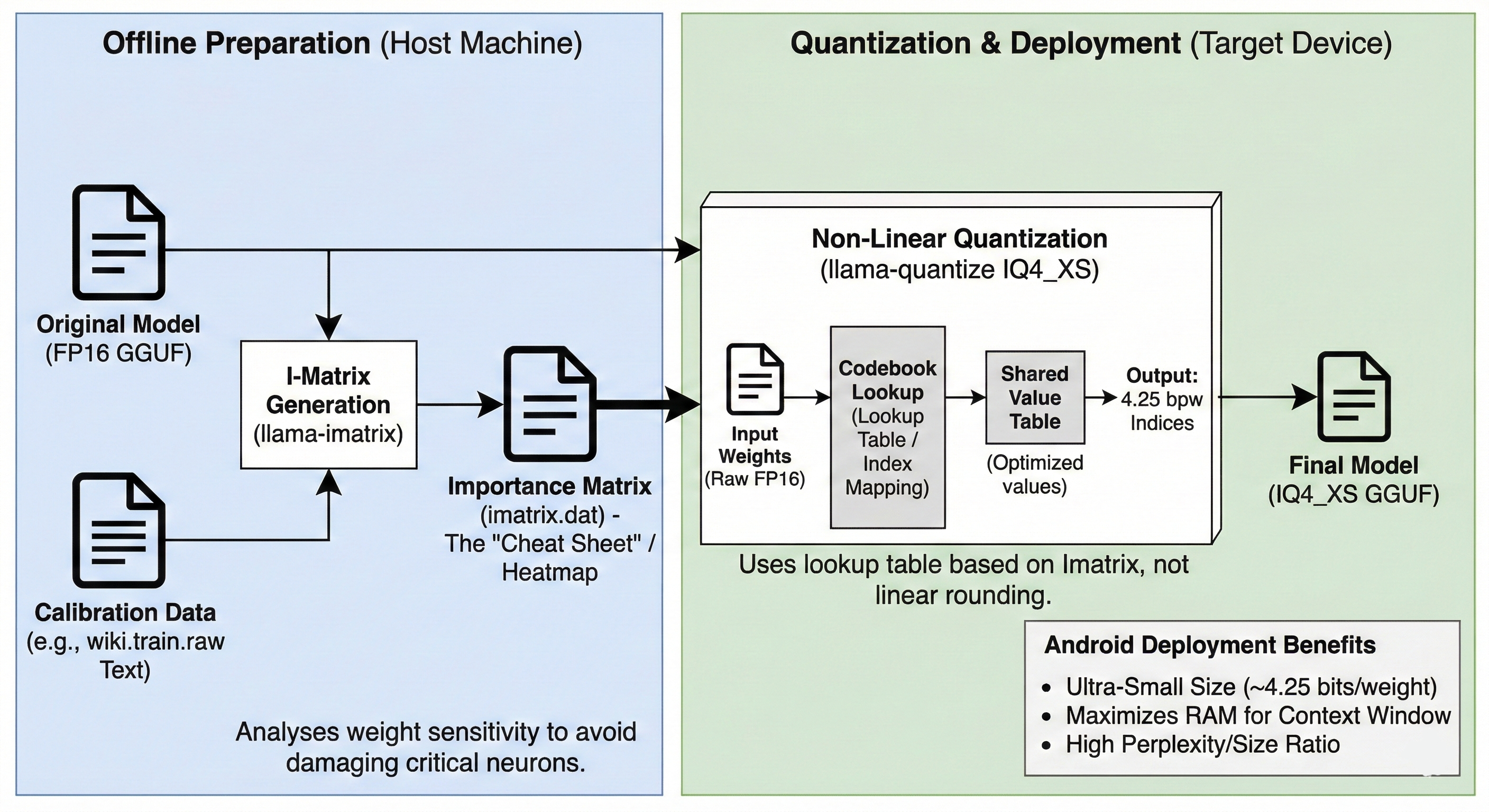}
    \caption{\texttt{IQ4\_XS}: importance-aware 4-bit quantization using calibration data and codebook-based reconstruction.}
    \label{fig:iq4xs}
\end{figure}

The second format evaluated is \texttt{IQ4\_XS}, an importance-aware quantization method inspired by activation-aware weight quantization techniques. Foundational research in this area~\cite{AWQ} shows that protecting a small fraction of salient weights, sometimes as little as one percent of the total parameter count, can reduce quantization error while still achieving high compression ratios. This quantization process begins with calibration data that is statistically similar to the original training distribution of the model. This data is passed through the network to calculate activation statistics, estimate importance metrics, and form a sensitivity map of the model weights. Instead of simply rounding weights to nearest fixed values, \texttt{IQ4\_XS} employs a \textbf{codebook-based approach}. The most important weights are assigned discrete values closely matching their original continuous representations to minimize accuracy loss, while less important weights undergo more aggressive compression. The model stores pointer indices directing to entries in a shared numerical codebook, which the device runtime references and reconstructs dynamically during inference. This methodology allocates precision where it is most needed and reduces the resident memory footprint. 
Figures~\ref{fig:q4km} and~\ref{fig:iq4xs} illustrate the structural differences between these two quantization strategies.

\subsection{\textbf{Experimental Setup and Orchestration}}
\label{subsec:setup_orchestration}

We designed the evaluation framework to reflect a realistic, unrooted consumer smartphone setting while maintaining automated run-to-run reproducibility suitable for empirical software engineering research.

The testing system comprises three main components. The host controller is a dedicated MacBook Air M4 that executes the Experiment-Runner~\cite{experiment-runner} orchestration framework. This controller defines experiments, deploys artifacts to the device, starts and stops hardware measurements, and collects telemetry logs. The target device and active inference engine is a Samsung Galaxy S25 Ultra running the Android 16 operating system.

We selected this device for three reasons. First, it is representative of high-end Android hardware, making our findings relevant to the current class of flagship consumer smartphones. Second, during preliminary testing, the evaluated LLMs failed to run reliably on mid- and low-end devices due to insufficient RAM and thermal headroom, confirming that a high-end device is required for this parameter range. Third, the larger unified memory of a flagship device allowed us to evaluate a broader set of models – from 0.5B to 9B parameters – without triggering out-of-memory terminations, enabling a more complete comparison. This device executes the \texttt{llama.cpp} binary for CPU-only inference, allocating up to 8 processing threads to use the heterogeneous multi-core architecture of the mobile processor. Finally, the communication interface handles automated device control via Wireless Android Debug Bridge (ADB)~\cite{adb_google}. This wireless protocol is necessary for power profiling, as a physical USB connection would introduce charging currents from the host machine and \textbf{corrupt discharge-based energy measurements} recorded by the device.

\subsection{\textbf{Execution Pipeline and Reproducibility Controls}}
\label{subsec:pipeline}

\begin{figure}[t]
    \centering
    \scriptsize
    \begin{tikzpicture}[
        node distance=0.4cm and 0.5cm,
        process/.style={rectangle, draw, fill=blue!10, text width=3cm, align=center, rounded corners, minimum height=0.4cm},
        arrow/.style={thick, -{Stealth[length=1.5mm]}, rounded corners}
    ]
    \node (setup) [process, fill=orange!10] {
        Phase 1: One-Time Setup \\
        Configure experiment \\  Screen on \\  Background isolation \\  Deploy \& Warm-up
    };

    \node (startrun) [process, below=0.7cm of setup] {1) Start Run \\ Clear logcat};
    \node (startmeasure) [process, below=of startrun] {2) Start Measurement \\ Start logger (100ms) + 2s spin-up};
    \node (inference) [process, below=of startmeasure] {3) Run Inference \\ \texttt{llama.cpp} (local output)};
    \node (stopmeasure) [process, below=of inference] {4) Stop Measurement \\ Stop logger};
    \node (collect) [process, below=of stopmeasure] {5) Collect \& Cleanup \\ Pull logs/outputs \\ Delete device artifacts};
    \node (cooldown) [process, fill=red!10, below=of collect] {6) Cool-Down \\ 200s pause};

    \draw [arrow] (setup) -- (startrun);
    \draw [arrow] (startrun) -- (startmeasure);
    \draw [arrow] (startmeasure) -- (inference);
    \draw [arrow] (inference) -- (stopmeasure);
    \draw [arrow] (stopmeasure) -- (collect);
    \draw [arrow] (collect) -- (cooldown);
    \draw [arrow] (cooldown.east) -- ++(1.4,0) |- (startrun.east) node[pos=0.20, left] {next run};
    \end{tikzpicture}
    \caption{Controlled execution pipeline. Setup is executed once; measurement and inference are repeated per run with run-level isolation and thermal management.}
    \label{fig:pipeline}
\end{figure}
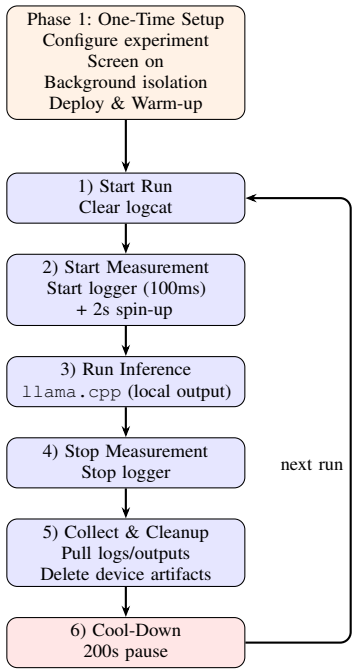

We structured the methodology as a controlled pipeline consisting of a one-time global setup phase, a repeated experimental run loop, and a final environmental restoration phase (Fig.~\ref{fig:pipeline}). This design ensures measurement isolation across consecutive runs and mitigates environmental confounds, such as background OS activity and thermal carry-over from previous computations.

\textbf{\emph{Pre-experiment configuration:}} Before collecting empirical data, we standardize the device state during the one-time setup phase. We first define a global configuration table specifying the model name, the targeted task type, and 30 repetitions to support statistically stable results. The smartphone screen is programmatically \textbf{kept on and locked at minimum brightness}. This is required because allowing the display to turn off triggers Android OS idling behaviors that reduce CPU clock frequencies and limit processing throughput to save battery~\cite{hoque2015understanding}, thereby skewing latency and energy metrics. Furthermore, background activities, such as periodic Google Play services, automated bug reporting, and application maintenance tasks, are isolated. This prevents unrelated CPU usage spikes and secondary battery drains from interfering with our baseline measurements. Finally, we clear the system \texttt{logcat} to establish a clean logging environment.

\textbf{\emph{Model deployment and warm-up:}} We automatically push all required LLM files and the \texttt{llama.cpp} execution engine to the local device storage. An initial warm-up phase is then executed. This preliminary inference run prepares device memory, populates system caches, stabilizes CPU frequency scaling governors, and reduces the impact of cold-start penalties on power behavior.

\textbf{\emph{Experimental loop:}} Following the one-time global setup, the orchestration framework enters an execution loop for each defined iteration. Run-level log isolation is enforced by clearing the \texttt{logcat} buffer again at the start of each run. If system logs were only cleared at the beginning of the experiment, previous run data would accumulate, making it difficult to parse, separate, and analyze individual trials without data contamination. The on-device battery logger is then initialized with a sampling rate of 100 milliseconds. This is followed by a spin-up period of one to two seconds, which allows the monitoring system to stabilize its telemetry reporting before inference begins.

During the interaction stage, the language model is invoked using model-specific instruction templates. This is necessary because each model has been fine-tuned on different datasets and expects a specific prompt structure to function properly. The generated textual outputs are written to the local storage of the device. To keep each run isolated from the next, the host machine pulls these output files over the wireless network and deletes them from the smartphone. Without this cleanup step, leftover text files and KV cache artifacts could consume device storage, potentially interfering with filesystem performance and subsequent telemetry analysis.

\textbf{\emph{Post-run collection and restoration:}} Once generation concludes, the battery logger and active system profilers are terminated. All recorded time-series logs containing device temperature, current, voltage, and synchronized timestamps are pulled back to the host controller. In the final data population stage, raw performance metrics and textual responses are consolidated into a structured CSV file for downstream evaluation. After all scheduled runs complete, an after-experiment phase restores the smartphone to its normal operating state by terminating lingering monitoring services, clearing residual experiment logs, and resetting the screen timeout and system configurations to their original defaults.

\emph{Reproducibility parameters.} We fixed decoding and runtime parameters to ensure comparability across trials. The maximum output generation length is bounded to 100 tokens, as variable generation lengths would skew total energy measurements and make cross-model comparisons invalid. The context window size is set to 512 tokens, generation temperature is fixed at 0 to force deterministic greedy decoding, and the inference process uses 8 CPU threads. To manage thermal dynamics and prevent heat accumulation, a \textbf{200-second cool-down period} is enforced between runs, allowing the passive cooling of the smartphone chassis to return the processor to a baseline temperature. Furthermore, the internal battery range is maintained between 80 percent and 100 percent charge, evaluating one model per testing round. This prevents voltage drops that occur at lower charge levels, such as the shift from 4.4 V at full charge to lower thresholds, ensuring models are evaluated under comparable electrical conditions. The input prompt text is fixed, though the exact token count varies between 93 and 119 tokens across models due to vocabulary and tokenizer differences.

\subsection{\textbf{Energy Measurement and Post-Processing}}
\label{subsec:energy_measurement}

Accurate power profiling on consumer smartphones presents methodological challenges due to restricted telemetry access and tightly integrated hardware. Three paradigms exist, each with trade-offs between measurement fidelity and realistic device behavior. 
External hardware monitors such as Monsoon~\cite{monsoon_hvpm} offer circuit-level precision but require invasive battery bypass, altering the device's thermal characteristics and undermining realistic operating conditions. ADB-based extraction is less invasive but blocked by kernel-level security policies on modern Android devices, making it impractical without root access~\cite{li2022power}. We intentionally avoided rooting to preserve realistic stock software conditions.

\begin{figure}[t]
    \centering
    \includegraphics[width=\linewidth]{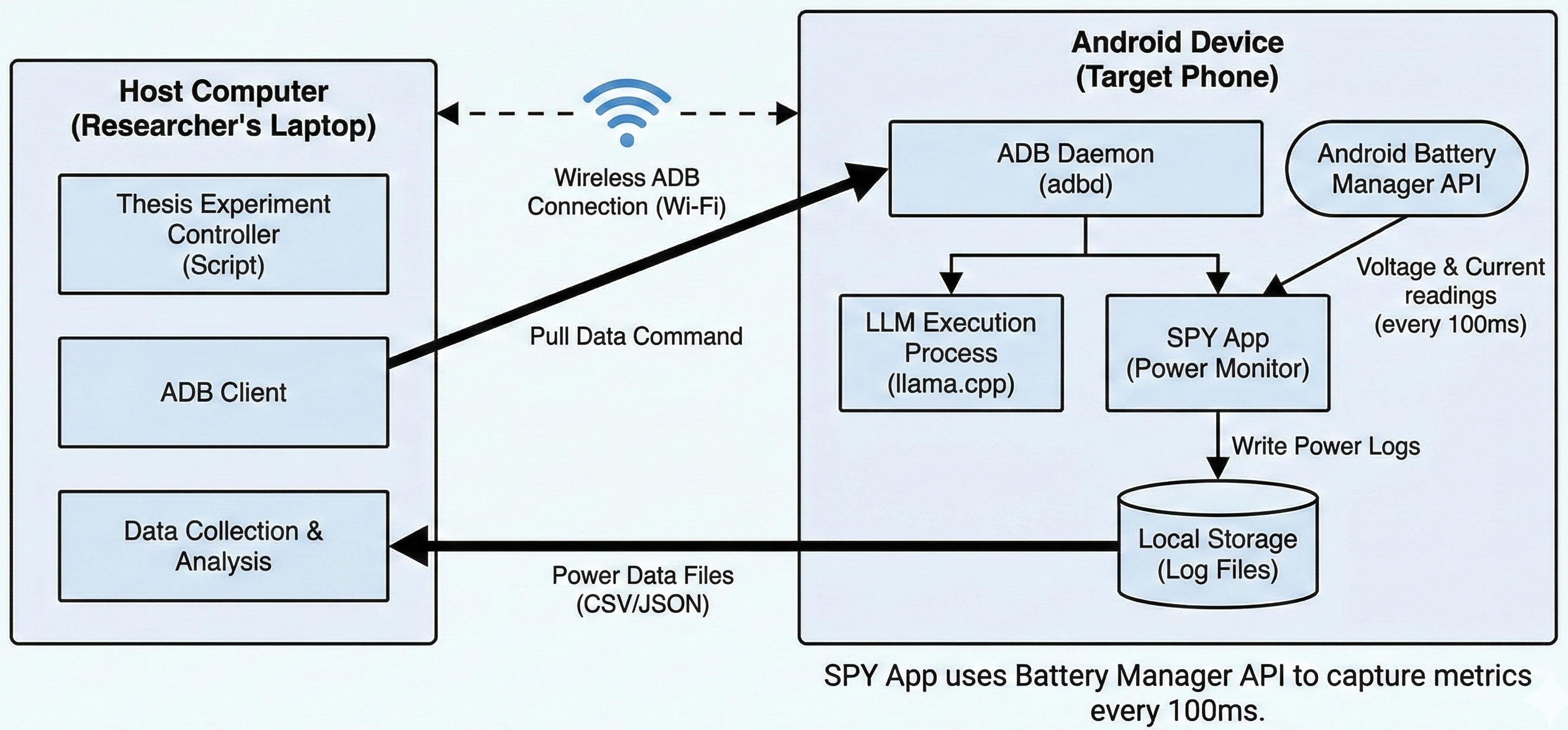}
    \caption{Measurement workflow. A user-space monitoring app logs voltage, current, and 
    temperature at fixed intervals while \texttt{llama.cpp} runs on-device. Logs are pulled 
    to the host after each run.}
    \label{fig:monitoring}
\end{figure}

To overcome these limitations, we adopted an on-device background monitoring application leveraging the native Android BatteryManager API~\cite{BatteryManager_API,battery-manager}. We developed a dedicated logging service installed directly on the target smartphone as a standard internal application; the Android security framework permits this telemetry without requiring root access. During the execution pipeline, the orchestration framework triggers this application to measure energy consumption alongside inference. This method \textbf{preserves realistic deployment conditions}. This application periodically logs battery voltage, current, temperature, and Unix timestamps at a \textbf{100-millisecond} sampling interval. A brief spin-up period of one to two seconds allows the monitor to stabilize before inference begins. Logs are retrieved by the host controller after each run completes to avoid interfering with the measurement window (Fig.~\ref{fig:monitoring}).

Following data collection, we isolate the energy attributable to the language model. Instantaneous gross power is computed by multiplying the sampled voltage and current at each time step. 
Because an idle smartphone draws background power, we perform a \textbf{baseline subtraction}. We recorded the device in an idle state for two hours under matched environmental conditions, with the screen on at minimum brightness and no foreground workload, and subtract the mean baseline power $P_{\text{baseline}}$ from the gross signal to isolate net inference power. Finally, total energy per run is computed by integrating net power using trapezoidal integration. With $N$ samples at timestamps $\{t_n\}$, the total Joules consumed are:
\begin{equation}
    E = \sum_{n=0}^{N-1}\frac{P_{\text{net}}(t_n)+P_{\text{net}}(t_{n+1})}{2}
    \cdot (t_{n+1}-t_n). \nonumber
\end{equation}

\subsection{\textbf{Evaluation Metrics}}
\label{subsec:metrics}

We report results across multiple dimensions, including hardware performance, resource usage, energy efficiency, and semantic output quality.

Performance metrics include prefill speed measured in tokens per second, which represents prompt processing throughput. Because prefill processes the input sequence in parallel, it primarily reflects the computational capacity of the device. We also measure generation speed in tokens per second for autoregressive decoding throughput. Because decoding is sequential, it is strongly bound by the memory bandwidth of the device. Additionally, we track Time to First Token to measure startup latency until the initial token appears, and end-to-end latency for the combined duration of the prefill and generation phases. Efficiency metrics comprise total integrated energy measured in Joules, energy consumed per generated token, and peak memory measured in gigabytes, encompassing model weights, the dynamic KV cache, and runtime buffers.

Output quality is assessed through two complementary tracks using both reference-based and reference-free evaluation methodologies. For reference-based evaluation, we use BERTScore~\cite{bert_score}. Instead of relying on exact word overlap metrics, BERTScore uses contextual embeddings to measure semantic similarity between the generated response of a model and a human-authored gold-standard reference summary. This provides a robust evaluation of semantic quality, particularly for summarization tasks where vocabulary and sentence structure may vary while the underlying meaning remains consistent.

However, during evaluation, we observed a recurring limitation of reference-based metrics when applied across diverse model sizes. Smaller models frequently achieved higher semantic similarity scores than larger counterparts, even when human inspection revealed summaries of lower quality and readability. This occurred because smaller models tended to copy large, contiguous portions of the source text with limited restructuring or abstraction. Because BERTScore measures semantic similarity, outputs that closely resemble the original source text can receive high scores. In contrast, larger models were more likely to paraphrase, logically restructure sentences, and produce more natural summaries. Although these outputs were often better in readability and abstraction, they occasionally received lower BERTScore values because they diverged from the wording of the reference material. This revealed that the metric favored extractive behavior over abstractive summarization, \textbf{biasing the evaluation toward less capable models}~\cite{fabbri2021summeval}. 
Consequently, to address this metric bias, we relied on reference-free qualitative evaluation to capture coherence, faithfulness, and overall summary quality beyond surface-level similarity. We implemented an LLM-based judging mechanism inspired by the G-Eval method~\cite{G-Eval}. In this setup, a capable language model independently assesses the quality of each generated response according to predefined criteria without relying on a gold reference. The evaluation pipeline scores generated responses based on faithfulness, to ensure the summary reflects the factual source content; relevance, to confirm the response focuses on critical information rather than trivial details; and coherence, to verify that the text is logically structured and flows naturally. These dimensions are combined into an overall quality score reported as the final evaluation metric~\cite{G-Eval}.

We statistically analyze per-run measurements based on the 30 repetitions recorded per condition. Normality across the dataset is assessed using the Shapiro-Wilk test. Because power profiling and latency distributions on mobile operating systems are consistently non-normal, we employ \textbf{non-parametric tests}. We use the Friedman test for omnibus comparisons across all models, and paired Wilcoxon signed-rank tests combined with the Holm-Bonferroni correction for post-hoc multi-model and quantization rankings.

\section{Results}
\label{sec:results}


To make the connection between the empirical results and the research questions explicit, we organize the findings as follows. Section~\ref{subsec:results_RQ1} addresses RQ1 by examining whether on-device LLM deployment is practically feasible in terms of throughput, latency, energy consumption, and user-facing responsiveness. Section~\ref{subsec:results_RQ2} addresses RQ2 by isolating the effects of the two 4-bit quantization formats on memory, speed, latency, and energy. Section~\ref{subsec:results_RQ3} addresses RQ3 by analyzing how model scale and architecture, including dense and Mixture-of-Experts models, affect performance and efficiency on mobile hardware. Section~\ref{subsec:results_implications} summarizes the practical deployment implications emerging from the three research questions.

As for the statistical analysis, we first assess the assumption of normality for all five key performance and efficiency metrics using the Shapiro-Wilk test, detailed in Table~\ref{tab:shapiro_wilk}. As anticipated in Section \ref{sec:methodology}, the p-value is significantly less than the 0.05 for every evaluated metric, rejecting the null hypothesis of normally distributed samples. Consequently, we rely exclusively on non-parametric tests for all subsequent analyses. 

\begin{table}[ht]
\centering
\caption{Shapiro--Wilk test results.}
\label{tab:shapiro_wilk}
\small
\renewcommand{\arraystretch}{1.3}
\begin{tabular}{lcc}
\hline
\rowcolor{charcoal}
\textcolor{white}{\textbf{Metric}} & \textcolor{white}{\textbf{W-Statistic}} & \textcolor{white}{\textbf{p-value}} \\
\hline
Generation Speed        & 0.7375 & 3.1716e-19 \\
\rowcolor{lightgray!40}
Energy Consumption      & 0.8800 &  7.7721e-13 \\
Memory Usage            & 0.8885 & 2.6160e-12 \\
\rowcolor{lightgray!40}
Latency                 & 0.8540 & 2.6560e-14 \\
Time to First Token     & 0.8574 & 4.0479e-14 \\
\hline
\end{tabular}
\end{table}


\subsection{RQ1: Feasibility of Efficient On-Device LLM Deployment}
\label{subsec:results_RQ1}

\subsubsection{\textbf{Throughput}}
\label{subsec:results_perf}

We measure performance across two distinct operational phases of LLM inference: prompt processing throughput, referred to as the prefill phase, and autoregressive decoding throughput, referred to as the generation phase. Figure~\ref{fig:speed-analysis} demonstrates a large and consistent performance separation between these two computational phases across all evaluated neural architectures and both quantization schemes.

\begin{figure*}[t]
    \centering
    \begin{minipage}{0.48\textwidth}
        \centering
        \begin{tikzpicture}
\begin{axis}[
    ybar,
    area legend,
    bar width=7pt,
    width=\linewidth,
    height=6cm,
    enlarge x limits=0.2,
    ylabel={Prefill Speed (tokens/s)},
    font=\footnotesize,
    ymin=0,
    ymax=220,
    ytick distance=50,
    xtick=data,
    table/col sep=comma,
    xticklabels from table={data/Q4_K_M.csv}{model_name},
    x tick label style={rotate=45, anchor=east, font=\small},
    legend style={at={(0.97,0.95)}, anchor=north east, legend columns=1, font=\scriptsize},
    ymajorgrids=true,
    grid style=dashed,
]

\addplot[
    color=blue, 
    fill=blue!50,
    error bars/.cd,
    y dir=both,
    y explicit
] table[
    x expr=\coordindex,
    y=prompt_prefill_speed,
    y error=prefill_IQR,
    col sep=comma
] {data/Q4_K_M.csv};

\addplot[
    color=blue, 
    fill=blue!20,
    error bars/.cd,
    y dir=both,
    y explicit
] table[
    x expr=\coordindex,
    y=prompt_prefill_speed,
    y error=prefill_IQR,
    col sep=comma
] {data/IQ4_XS.csv};

\legend{Q4\_K\_M, IQ4\_XS}

\end{axis}
\end{tikzpicture}
        \vspace{1mm}
        (a) Prefill Speed
    \end{minipage}
    \hfill
    \begin{minipage}{0.48\textwidth}
        \centering
        \begin{tikzpicture}
\begin{axis}[
    ybar,
    area legend,
    bar width=7pt,
    width=\linewidth,
    height=6cm,
    enlarge x limits=0.2,
    ylabel={Generation Speed (tokens/s)},
    font=\footnotesize,
    ymin=0,
    ymax=60,
    ytick distance=10,
    xtick=data,
    table/col sep=comma,
    xticklabels from table={data/Q4_K_M.csv}{model_name},
    x tick label style={rotate=45, anchor=east, font=\small},
    legend style={at={(0.97,0.95)}, anchor=north east, legend columns=1, font=\scriptsize},
    ymajorgrids=true,
    grid style=dashed,
]

\addplot[
    color=orange, 
    fill=orange!50,
    error bars/.cd,
    y dir=both,
    y explicit
] table[
    x expr=\coordindex,
    y=generation_decoder_speed,
    y error=generation_IQR, 
    col sep=comma
] {data/Q4_K_M.csv};

\addplot[
    color=orange, 
    fill=orange!20,
    error bars/.cd,
    y dir=both,
    y explicit
] table[
    x expr=\coordindex,
    y=generation_decoder_speed,
    y error=generation_IQR, 
    col sep=comma
] {data/IQ4_XS.csv};

\legend{Q4\_K\_M, IQ4\_XS}

\end{axis}
\end{tikzpicture}
        \vspace{1mm}
        (b) Generation Speed
    \end{minipage}
    \caption{Throughput across models and quantization schemes. Prefill benefits from parallel processing of the prompt, whereas generation is slower due to autoregressive decoding and higher per-token overhead.}
    \label{fig:speed-analysis}
\end{figure*}

\textbf{\emph{Prefill versus generation dynamics:}} Across all hardware and software configurations, prefill throughput exceeds generation throughput. This discrepancy is deeply rooted in the underlying mathematics of the Transformer architecture. The prefill phase processes the entire input sequence simultaneously, allowing the inference engine to utilize optimized, parallel matrix-matrix multiplications to efficiently populate the initial Key-Value cache~\cite{kwon2023efficient}. In contrast, the generation phase is inherently sequential. Each newly generated token depends on all previously generated tokens, forcing the system to rely on memory-bound matrix-vector multiplications. This sequential dependency prevents parallel execution across the sequence length, pushing the processor into the memory wall, a phenomenon where the \textbf{speed of data transfer from RAM dictates performance} rather than the arithmetic speed of the CPU~\cite{wulf1995hitting}. This increases the end-to-end runtime and establishes the generation phase as the dominant bottleneck and the primary contributor to total inference time for any moderately long output sequence.

\textbf{\emph{Scaling with model size:}} As anticipated by theoretical computational complexity bounds and established scaling laws~\cite{kaplan2020scaling}, smaller models consistently achieve much higher throughput, while larger multi-billion parameter models execute slower under both 4-bit quantization formats. This downward trend directly reflects the increased compute requirements and memory traffic associated with larger parameter counts and larger intermediate attention states. Every generated token requires loading the entirety of the model weights from the system RAM into the limited L1 and L2 caches of the mobile CPU, creating a severe memory bandwidth bottleneck that mobile systems, constrained by low-power LPDDR memory interfaces, are poorly equipped to handle~\cite{david2021tensorflow}.

\subsubsection{\textbf{Latency}}
\label{subsec:results_latency}

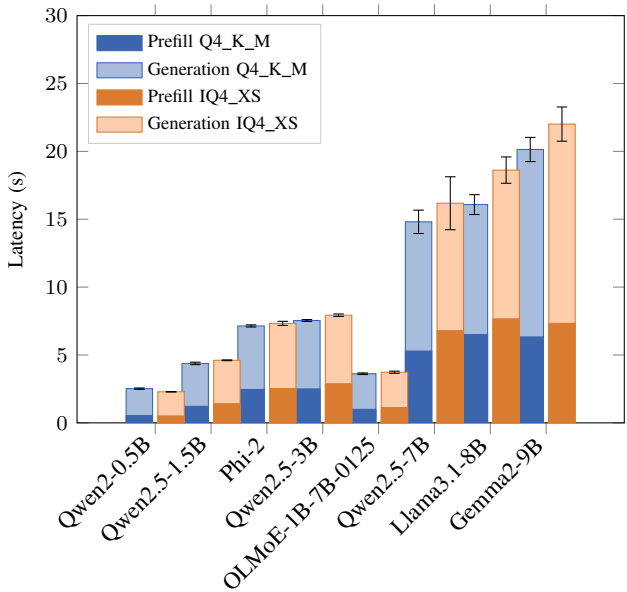
\begin{figure}[t]
\centering
\begin{tikzpicture}
\begin{axis}[
    ybar, 
    bar width=10pt,
    width=\linewidth,
    height=7cm,
    enlarge x limits=0.2,
    ylabel={Latency (s)},
    font=\footnotesize,
    ymin=0,
    ymax=30, 
    ytick distance=5,
    xtick=data,
    table/col sep=comma,
    xticklabels from table={data/Q4_K_M.csv}{model_name},
    x tick label style={rotate=45, anchor=east, font=\small},
    legend style={
        at={(0.02,0.98)},
        anchor=north west,
        legend columns=1,
        font=\scriptsize,
        draw=black!20,
        fill=white,
        cells={anchor=west}
    },
]

\definecolor{Q4Dark}{RGB}{59, 101, 175}   
\definecolor{Q4Light}{RGB}{168, 189, 219} 
\definecolor{IQ4Dark}{RGB}{218, 124, 48}  
\definecolor{IQ4Light}{RGB}{253, 205, 172}

\addplot[
    forget plot,
    color=Q4Dark, 
    fill=Q4Light, 
    bar shift=-6pt,
    error bars/.cd, y dir=both, y explicit, error bar style={black}
] table[
    x expr=\coordindex,
    y expr=\thisrow{prefill_latency} + \thisrow{generation_latency},
    y error=inference_IQR, 
    col sep=comma
] {data/Q4_K_M.csv};

\addplot[
    forget plot, 
    color=Q4Dark, 
    fill=Q4Dark, 
    bar shift=-6pt,
] table[
    x expr=\coordindex,
    y=prefill_latency,
    col sep=comma
] {data/Q4_K_M.csv};

\addplot[
    forget plot,
    color=IQ4Dark, 
    fill=IQ4Light, 
    bar shift=6pt,
    error bars/.cd, y dir=both, y explicit, error bar style={black}
] table[
    x expr=\coordindex,
    y expr=\thisrow{prefill_latency} + \thisrow{generation_latency},
    y error=inference_IQR, 
    col sep=comma
] {data/IQ4_XS.csv};

\addplot[
    forget plot, 
    color=IQ4Dark, 
    fill=IQ4Dark, 
    bar shift=6pt,
] table[
    x expr=\coordindex,
    y=prefill_latency,
    col sep=comma
] {data/IQ4_XS.csv};

\addlegendimage{ybar, area legend, color=Q4Dark, fill=Q4Dark}
\addlegendentry{Prefill Q4\_K\_M}

\addlegendimage{ybar, area legend, color=Q4Dark, fill=Q4Light}
\addlegendentry{Generation Q4\_K\_M}

\addlegendimage{ybar, area legend, color=IQ4Dark, fill=IQ4Dark}
\addlegendentry{Prefill IQ4\_XS}

\addlegendimage{ybar, area legend, color=IQ4Dark, fill=IQ4Light}
\addlegendentry{Generation IQ4\_XS}

\end{axis}
\end{tikzpicture}
\caption{Inference latency breakdown.}
\label{fig:inference-latency}
\end{figure}

Figure~\ref{fig:inference-latency} visualizes the total inference latency, decomposed into the initial prefill time and the subsequent generation time. Each bar represents the median end-to-end latency required to produce a sequence of up to 100 output tokens.

\textbf{\emph{Latency breakdown and user experience:}} Absolute latency increases substantially and non-linearly with model size. The graphical breakdown illustrates that the autoregressive generation phase dominates the total latency budget, especially for the larger models, which aligns with the sequential, memory-bound nature of the decoding process. This has direct implications for practical human deployment. Established guidelines in human-computer interaction dictate that interaction latencies exceeding 200 milliseconds rapidly become perceptible to human users, and systemic delays extending beyond two to three seconds disrupt natural conversational flow and cause cognitive friction~\cite{miller1968response,nielsen1993usability}. As results demonstrate, several of the larger models, such as the 8B and 9B variants, exhibit end-to-end latencies extending well beyond 10 seconds per query. Consequently, while these models are technically feasible to execute on an unrooted mobile device, they are \textbf{practically unusable for real-time, interactive digital assistant applications} where immediate responsiveness is the paramount design requirement.

\subsubsection{\textbf{Total Energy and Practical Usability}}
\label{subsec:results_energy_feasibility}
Beyond raw execution speed, practical feasibility on mobile hardware also depends on the energy required to complete an inference request. A model that can technically run on the device may still be unsuitable for real-world deployment if each interaction consumes excessive battery or if the required 
computation increases heat and trigger throttling. The latency analysis is complemented with total energy consumption and energy-per-token measurements.

\begin{figure}[t]
\centering

\pgfplotstableread[col sep=comma]{data/Q4_K_M.csv}\datatable

\begin{tikzpicture}
    \pgfplotsset{
        width=0.85\linewidth,
        height=5cm,
        scale only axis,
        xtick=data,
        xticklabels from table={\datatable}{model_name},
        x tick label style={rotate=45, anchor=east, font=\small},
    }

    \begin{axis}[
        axis y line*=left,
        ybar,
        bar width=7pt,
        ylabel={Time to First Token (s)},
        font=\scriptsize,
        ymin=0,
        enlarge x limits=0.15,
        grid=none,
        legend style={
            at={(0.05,0.95)},
            anchor=north west,         
            legend columns=1, 
        },
    ]
    
    \addplot[
        draw=none, 
        fill=blue, 
        fill opacity=0.3
    ] table[
        x expr=\coordindex,
        y=time_to_first_token, 
        col sep=comma
    ] {data/Q4_K_M.csv};
    \addlegendentry{Q4\_K\_M (Time)}

    \addplot[
        draw=none, 
        fill=red, 
        fill opacity=0.3
    ] table[
        x expr=\coordindex,
        y=time_to_first_token, 
        col sep=comma
    ] {data/IQ4_XS.csv};
    \addlegendentry{IQ4\_XS (Time)}

    \addlegendimage{line legend, color=blue, mark=*, mark size=2pt, thick}
    \addlegendentry{Q4\_K\_M (Energy)}
    
    \addlegendimage{line legend, color=red, mark=square*, mark size=2pt, thick}
    \addlegendentry{IQ4\_XS (Energy)}
    
    \end{axis}

    \begin{axis}[
        axis y line*=right,
        axis x line=none,
        ylabel={Energy per Token (Joules)},
        font=\scriptsize,
        ymin=0, ymax=4,
        enlarge x limits=0.15,
        grid=none,
        grid style=dashed,
    ]

    \addplot[
        color=blue, 
        mark=*, 
        mark size=2pt, 
        thick
    ] table[
        x expr=\coordindex,
        y=energy_per_token, 
        col sep=comma
    ] {data/Q4_K_M.csv};

    \addplot[
        color=red, 
        mark=square*, 
        mark size=2pt, 
        thick
    ] table[
        x expr=\coordindex,
        y=energy_per_token, 
        col sep=comma
    ] {data/IQ4_XS.csv};

    \end{axis}
\end{tikzpicture}
\caption{Comparison of Time to First Token (Bars, Left Axis) and Energy per Token (Lines, Right Axis).}
\label{fig:energy-per-token}
\end{figure}
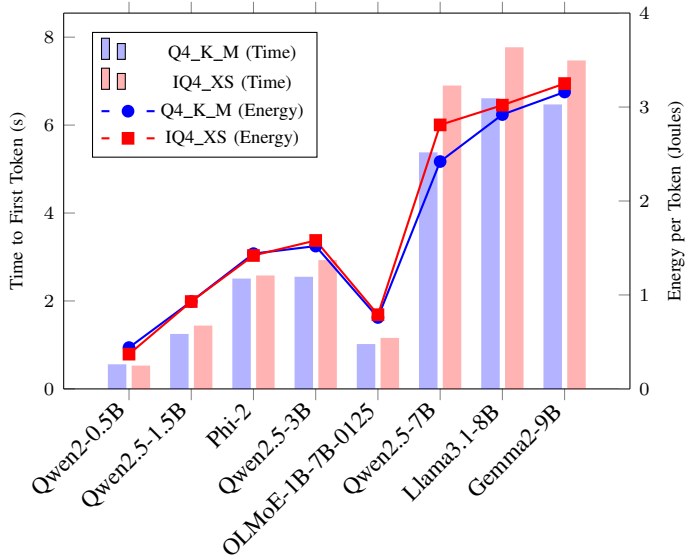

Hereafter, we evaluate both the initial startup energy cost, which is closely correlated with Time to First Token, and the sustained energy efficiency, quantified as physical Joules consumed per generated token. Figure~\ref{fig:energy-per-token} maps the relationship between initial latency and sustained energy cost across the entire spectrum of models and both quantization schemes.

Larger neural architectures exhibit both higher startup latencies and elevated energy costs per token. For example, the Meta-Llama-3.1-8B configuration demonstrates severe startup latency and high energy cost per token when compared to the smaller models. Similarly, the Gemma-2-9B model exhibits energy costs that are consistent with its computational complexity, parameter count, and memory bandwidth demands. In contrast, the smaller models, notably Qwen2-0.5B, manage to produce the first token almost instantaneously and consume a fraction of the energy per token, positioning them as viable candidates for latency-sensitive and battery-constrained mobile applications.

\begin{tcolorbox}[
    colback=paleblue,
    colframe=black,
    boxrule=0.5pt,
    left=1pt, right=1pt, top=1pt, bottom=1pt,
    ]
\faSearch\ \textbf{Answer to RQ$_1$.} On-device LLM deployment is technically feasible on a flagship smartphone across the evaluated 0.5B--9B range, since all models ran successfully without hardware failure or out-of-memory errors. However, practical feasibility is limited to smaller and mid-sized models. Models up to approximately 3B parameters offer acceptable generation speed and latency for interactive use, whereas larger dense models remain technically executable but are unsuitable for real-time assistant scenarios because of high latency and energy cost.
\end{tcolorbox}


\subsection{RQ2: Impact of Quantization on On-Device LLMs}
\label{subsec:results_RQ2}

\textbf{\emph{The limited impact of quantization format:}} Across the vast majority of evaluated models, the measured energy and speed differences between the standard \texttt{Q4\_K\_M} format and the complex \texttt{IQ4\_XS} format are modest, especially when compared to the disparities between different model families and overall parameter scales. This suggests a critical architectural insight. Under CPU-only inference on contemporary mobile hardware, the baseline model architecture and the overall volume of active arithmetic compute dominate physical energy consumption far more than the specific algorithmic choice between these two distinct 4-bit compression formats. The computational overhead required to unpack and reconstruct the codebook indices in \texttt{IQ4\_XS} introduces significant branching and cache lookup penalties~\cite{han2015deep}. These processor inefficiencies appear to \textbf{offset the theoretical energy savings} gained by moving fewer total bytes across the system bus from RAM.

\begin{figure*}[htbp]
    \centering
    \includegraphics[width=\linewidth]{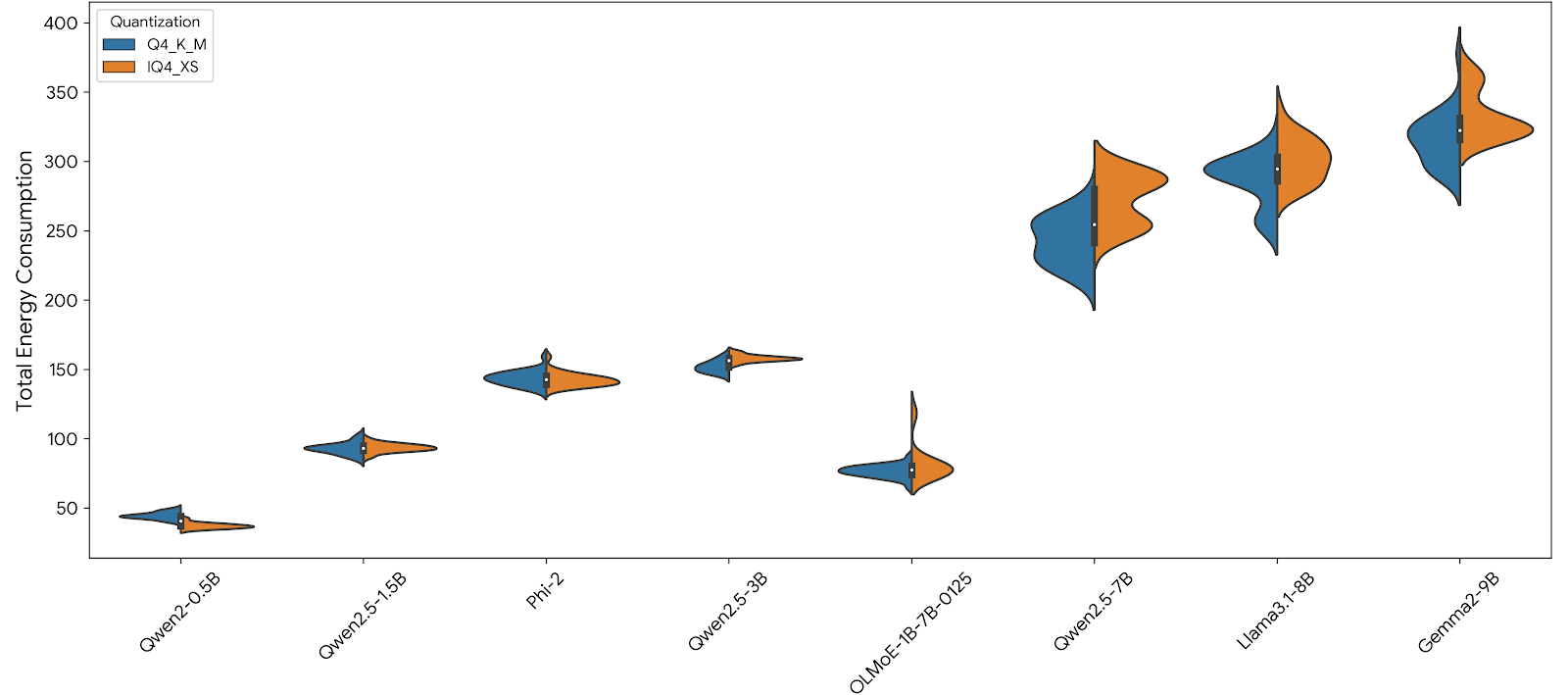}
    \caption{Total energy per run across models and quantization schemes. Distributions summarize 30 repetitions per configuration.}
    \label{fig:violin_energy}
\end{figure*}

\textbf{\emph{Energy distributions and thermal stability:}} Figure~\ref{fig:violin_energy} visualizes the complete per-run energy distributions using violin plots. These distributions reinforce the scaling trends for dense models and highlight the MoE deviation, showing OLMoE-1B-7B clustering tightly alongside the 1.5B and 3B dense models in absolute energy draw, despite its large resident weight capacity. Across all evaluated models, the two quantization variants produce broadly similar distribution shapes and internal variances. This variance spread indicates that while thermal throttling and background operating system scheduling introduce some expected run-to-run noise over 30 repetitions, the aggregate physical energy behavior of the device remains stable and bounded.

\textbf{\emph{Quantization effects and statistical reality:}} To quantify the impact of the selected quantization formats, we compare the paired \texttt{Q4\_K\_M} and \texttt{IQ4\_XS} runs using the Wilcoxon signed-rank test. Table~\ref{tab:wilcoxon_quantization} summarizes the results. While the codebook-based \texttt{IQ4\_XS} achieves a statistical victory in minimizing peak memory usage, the structurally simpler \texttt{Q4\_K\_M} achieves significant victories across all critical dynamic operational metrics, including generation speed, overall latency, and total physical energy consumption.

\begin{table}[ht]
\centering
\caption{Wilcoxon signed-rank test results for quantization comparison.}
\label{tab:wilcoxon_quantization}
\small
\renewcommand{\arraystretch}{1.3}
\begin{tabular}{lccc}
\hline
\rowcolor{charcoal}
\textcolor{white}{\textbf{Metric}} & \textcolor{white}{\textbf{W}} & \textcolor{white}{\textbf{p-value}} & \textcolor{white}{\textbf{Winner}} \\
\hline
Generation Speed     & 11179 & 0.00423  & \texttt{Q4\_K\_M} \\
\rowcolor{lightgray!40}
Energy Consumption   & 8967  & 3.36e-07 & \texttt{Q4\_K\_M} \\
Memory Usage         & 0     & 2.82e-41 & \texttt{IQ4\_XS} \\
\rowcolor{lightgray!40}
Latency              & 2966  & 1.32e-26 & \texttt{Q4\_K\_M} \\
Time to First Token  & 899   & 2.24e-36 & \texttt{Q4\_K\_M} \\
\hline
\end{tabular}
\end{table}

\begin{tcolorbox}[
    colback=paleblue,
    colframe=black,
    boxrule=0.5pt,
    left=1pt, right=1pt, top=1pt, bottom=1pt,
    ]
\faSearch\ \textbf{Answer to RQ$_2$.} While \texttt{IQ4\_XS} reliably reduces peak memory, it confers no meaningful advantage in energy or speed. The runtime overhead of codebook reconstruction offsets its theoretical efficiency gains, and \texttt{Q4\_K\_M} wins on all dynamic operational metrics. Quantization format therefore matters when fitting a model into RAM, but the choice between these two 4-bit schemes has negligible consequences for battery life or generation throughput.
\end{tcolorbox}

\subsection{RQ3: Impact of Model Size and Architecture}
\label{subsec:results_RQ3}

\textbf{\emph{Dense versus Mixture-of-Experts architectures:}} We observe a distinct and favorable operational behavior for the sparse Mixture-of-Experts model, OLMoE-1B-7B. Although this specific architecture possesses a total parameter count nearing seven billion, the internal routing mechanism ensures that only a selected subset of neural experts is actively engaged per token~\cite{shazeer2017outrageously}. This dynamic, input-dependent routing \textbf{reduces the active computational burden per token} relative to standard dense models of a similar total size. Consequently, despite occupying the storage footprint of a large model, OLMoE-1B-7B produces a processing throughput that closely mirrors the performance of much smaller dense models in the one to two billion parameter class. This beneficial algorithmic sparsity effect appears consistently in both the prefill and generation throughput results.

\textbf{\emph{Energy versus quality trade-off:}} Figure~\ref{fig:all-tradeoffs}a visually demonstrates that the highest-quality configurations are linked to the most energy-consuming execution profiles. Gemma-2-9B achieves the best generative quality but demands a substantial energy budget for a portable device. However, we observe that raw parameter count alone does not strictly dictate the quality outcome. For instance, Qwen2.5-7B manages to achieve better quality while consuming measurably less total energy than Meta-Llama-3.1-8B. This indicates that base architectural design choices, tokenizer efficiency, and pre-training dataset quality influence the ultimate shape of the quality and efficiency frontier.

\textbf{\emph{Total energy and the role of architecture:}} Figure~\ref{fig:all-tradeoffs}b illustrates that total physical energy consumption increases almost exponentially with model scale for standard dense architectures. A single inference run utilizing Gemma-2-9B routinely exceeds 300 Joules, whereas the efficient Qwen2-0.5B operates near 50 Joules for the same summarization task. A notable deviation from this size-based scaling law is the MoE model. OLMoE-1B-7B consumes an amount of energy that is comparable to much smaller dense models, despite harboring a large total parameter count. This confirms that the reduced active compute per token achieved through sparse expert routing \textbf{translates into physical battery savings} on the hardware level.

\textbf{\emph{Speed versus quality trade-off:}} Figure~\ref{fig:all-tradeoffs}c reveals a strict Pareto frontier governing the general trend. Higher generation speed almost universally corresponds to lower cognitive quality as measured by the G-Eval framework. Faster models naturally tend to be much smaller in capacity, which improves physical efficiency metrics but limits their internal representations, hindering their ability to produce deep abstraction, logical restructuring, and high semantic coherence in complex summarization outputs. Conversely, massive models achieve superior semantic quality at the cost of reduced throughput and rapid battery drain.

\begin{figure*}[t]
    \centering
    \begin{minipage}{0.48\textwidth}
        \centering
        \begin{tikzpicture}

\def\getcolor#1{
  \ifcase#1 blue\or orange\or green!60!black\or red\or purple\or brown\or pink\or gray!60!black\fi%
}

\pgfplotstableread[col sep=comma]{data/Q4_K_M.csv}\tableQFour
\pgfplotstableread[col sep=comma]{data/IQ4_XS.csv}\tableIQFour

\pgfplotstablegetrowsof{\tableQFour}
\pgfmathtruncatemacro{\numrows}{\pgfplotsretval-1}

\begin{axis}[
    width=\linewidth,
    height=6cm,
    grid=major,
    xmin=0,
    xtick distance=50,
    ymin=0,
    ymax=1.0,
    ytick distance=0.2,
    xlabel={Total Energy Consumption (Joules)},
    ylabel={G-Eval Score},
    font=\footnotesize,
    enlarge x limits=0.08,
    enlarge y limits=0.10,
    legend columns=2,
    legend style={
        at={(0.98,0.98)},
        anchor=north east,
        font=\tiny,
        draw=none,
        fill=white,
        fill opacity=0.6,
        text opacity=1,
        cells={align=left, anchor=west},
        /tikz/column 2/.style={column sep=10pt}
    }
]

\pgfplotsforeachungrouped \i in {0,...,\numrows}{
    \pgfplotstablegetelem{\i}{model_name}\of{\tableQFour} \edef\modelname{\pgfplotsretval}
    
    \pgfplotstablegetelem{\i}{total_energy_consumption}\of{\tableQFour} \edef\xA{\pgfplotsretval}
    \pgfplotstablegetelem{\i}{Overall_G-Eval}\of{\tableQFour} \edef\yA{\pgfplotsretval}
    \pgfplotstablegetelem{\i}{total_energy_consumption}\of{\tableIQFour} \edef\xB{\pgfplotsretval}
    \pgfplotstablegetelem{\i}{Overall_G-Eval}\of{\tableIQFour} \edef\yB{\pgfplotsretval}

    \edef\temp{
        \noexpand\addplot[very thick, opacity=0.25, color=\getcolor{\i}, forget plot] coordinates { (\xA,\yA) (\xB,\yB) };
        \noexpand\addplot[only marks, opacity=0.25, mark=square*, mark size=2.6pt, color=\getcolor{\i}, forget plot] coordinates { (\xA,\yA) };
        \noexpand\addplot[only marks, opacity=0.7, mark=x, mark size=3.0pt, thick, color=\getcolor{\i}, forget plot] coordinates { (\xB,\yB) };
    }
    \temp
}

\end{axis}
\end{tikzpicture}
        (a) Energy vs.\ Quality
    \end{minipage}
    \hfill
    \begin{minipage}{0.48\textwidth}
        \centering
        \begin{tikzpicture}

\def\getcolor#1{
  \ifcase#1 blue\or orange\or green!60!black\or red\or purple\or brown\or pink\or gray!60!black\fi%
}

\pgfplotstableread[col sep=comma]{data/Q4_K_M.csv}\tableQFour
\pgfplotstableread[col sep=comma]{data/IQ4_XS.csv}\tableIQFour

\pgfplotstablegetrowsof{\tableQFour}
\pgfmathtruncatemacro{\numrows}{\pgfplotsretval-1}

\begin{axis}[
    width=\linewidth,
    height=6cm,
    grid=major,
    xmin=0,
    xtick distance=50,
    ymin=0,
    ytick distance=20,
    xlabel={Total Energy Consumption (Joules)},
    ylabel={Generation Speed (tokens/s)},
    font=\footnotesize,
    enlarge x limits=0.08,
    enlarge y limits=0.10,
    legend columns=2,
    legend style={
        at={(0.98,0.98)}, 
        anchor=north east,
        font=\tiny,
        draw=none,
        fill=none,
        fill opacity=0.6,
        text opacity=1,
        cells={align=left, anchor=west},
        /tikz/column 2/.style={column sep=10pt}
    }
]

\pgfplotsforeachungrouped \i in {0,...,\numrows}{

    \pgfplotstablegetelem{\i}{model_name}\of{\tableQFour} \edef\modelname{\pgfplotsretval}
    
    \pgfplotstablegetelem{\i}{total_energy_consumption}\of{\tableQFour} \edef\xA{\pgfplotsretval}
    \pgfplotstablegetelem{\i}{generation_decoder_speed}\of{\tableQFour} \edef\yA{\pgfplotsretval}
    \pgfplotstablegetelem{\i}{total_energy_consumption}\of{\tableIQFour} \edef\xB{\pgfplotsretval}
    \pgfplotstablegetelem{\i}{generation_decoder_speed}\of{\tableIQFour} \edef\yB{\pgfplotsretval}

    \edef\temp{
        \noexpand\addplot[very thick, opacity=0.25, color=\getcolor{\i}, forget plot] coordinates { (\xA,\yA) (\xB,\yB) };
        \noexpand\addplot[only marks, opacity=0.25, mark=square*, mark size=2.6pt, color=\getcolor{\i}, forget plot] coordinates { (\xA,\yA) };
        \noexpand\addplot[only marks, opacity=0.7, mark=x, mark size=3.0pt, thick, color=\getcolor{\i}, forget plot] coordinates { (\xB,\yB) };
    }
    \temp
}

\addlegendimage{empty legend}\addlegendentry{\textbf{Quantization}}
\addlegendimage{empty legend}\addlegendentry{\textbf{Model}}

\addlegendimage{only marks, mark=x, draw=gray, thick}\addlegendentry{IQ4\_XS}
\addlegendimage{only marks, mark=*, draw=gray!60!black, fill=gray!60!black}\addlegendentry{Gemma2-9B}

\addlegendimage{only marks, mark=square*, draw=gray, fill=gray}\addlegendentry{Q4\_K\_M}
\addlegendimage{only marks, mark=*, draw=pink, fill=pink}\addlegendentry{Llama3.1-8B}

\addlegendimage{empty legend}\addlegendentry{}
\addlegendimage{only marks, mark=*, draw=brown, fill=brown}\addlegendentry{Qwen2.5-7B}

\addlegendimage{empty legend}\addlegendentry{}
\addlegendimage{only marks, mark=*, draw=purple, fill=purple}\addlegendentry{OLMoE-1B-7B}

\addlegendimage{empty legend}\addlegendentry{}
\addlegendimage{only marks, mark=*, draw=red, fill=red}\addlegendentry{Qwen2.5-3B}

\addlegendimage{empty legend}\addlegendentry{}
\addlegendimage{only marks, mark=*, draw=green!60!black, fill=green!60!black}\addlegendentry{Phi-2}

\addlegendimage{empty legend}\addlegendentry{}
\addlegendimage{only marks, mark=*, draw=orange, fill=orange}\addlegendentry{Qwen2.5-1.5B}

\addlegendimage{empty legend}\addlegendentry{}
\addlegendimage{only marks, mark=*, draw=blue, fill=blue}\addlegendentry{Qwen2-0.5B}

\end{axis}
\end{tikzpicture}
        (b) Energy vs.\ Speed
    \end{minipage}

    \vspace{0.3cm}

    \begin{minipage}{0.48\textwidth}
        \centering
        \begin{tikzpicture}

\def\getcolor#1{
  \ifcase#1 blue\or orange\or green!60!black\or red\or purple\or brown\or pink\or gray!60!black\fi%
}

\pgfplotstableread[col sep=comma]{data/Q4_K_M.csv}\tableQFour
\pgfplotstableread[col sep=comma]{data/IQ4_XS.csv}\tableIQFour

\pgfplotstablegetrowsof{\tableQFour}
\pgfmathtruncatemacro{\numrows}{\pgfplotsretval-1}

\begin{axis}[
    width=\linewidth,
    height=6cm,
    grid=major,
    xmin=0,
    xtick distance=20,
    ymin=0,
    ymax=1.0,
    ytick distance=0.2,
    xlabel={Generation Speed (tokens/s)},
    ylabel={G-Eval Score},
    font=\footnotesize,
    enlarge x limits=0.08,
    enlarge y limits=0.10,
    legend columns=2,
    legend style={
        at={(0.98,0.98)},
        anchor=north east,
        font=\tiny,
        draw=none,
        fill=white,
        fill opacity=0.6,
        text opacity=1,
        cells={align=left, anchor=west},
        /tikz/column 2/.style={column sep=10pt}
    }
]

\pgfplotsforeachungrouped \i in {0,...,\numrows}{

    \pgfplotstablegetelem{\i}{model_name}\of{\tableQFour} \edef\modelname{\pgfplotsretval}
    
    \pgfplotstablegetelem{\i}{generation_decoder_speed}\of{\tableQFour} \edef\xA{\pgfplotsretval}
    \pgfplotstablegetelem{\i}{Overall_G-Eval}\of{\tableQFour} \edef\yA{\pgfplotsretval}
    \pgfplotstablegetelem{\i}{generation_decoder_speed}\of{\tableIQFour} \edef\xB{\pgfplotsretval}
    \pgfplotstablegetelem{\i}{Overall_G-Eval}\of{\tableIQFour} \edef\yB{\pgfplotsretval}

    \edef\temp{
        \noexpand\addplot[very thick, opacity=0.25, color=\getcolor{\i}, forget plot] coordinates { (\xA,\yA) (\xB,\yB) };
        \noexpand\addplot[only marks, opacity=0.25, mark=square*, mark size=2.6pt, color=\getcolor{\i}, forget plot] coordinates { (\xA,\yA) };
        \noexpand\addplot[only marks, opacity=0.7, mark=x, mark size=3.0pt, thick, color=\getcolor{\i}, forget plot] coordinates { (\xB,\yB) };
    }
    \temp
}

\end{axis}
\end{tikzpicture}
        (c) Speed vs.\ Quality
    \end{minipage}
    \hfill
    \begin{minipage}{0.48\textwidth}
        \centering
        \begin{tikzpicture}

\def\getcolor#1{
  \ifcase#1 blue\or orange\or green!60!black\or red\or purple\or brown\or pink\or gray!60!black\fi%
}

\pgfplotstableread[col sep=comma]{data/Q4_K_M.csv}\tableQFour
\pgfplotstableread[col sep=comma]{data/IQ4_XS.csv}\tableIQFour

\pgfplotstablegetrowsof{\tableQFour}
\pgfmathtruncatemacro{\numrows}{\pgfplotsretval-1}

\begin{axis}[
    width=\linewidth,
    height=6cm,
    grid=major,
    xmin=0,
    xtick distance=1,
    ymin=0,
    ymax=1.0,
    ytick distance=0.2,
    xlabel={Model Size (GB)},
    ylabel={G-Eval Score},
    font=\footnotesize,
    enlarge x limits=0.08,
    enlarge y limits=0.10,
    legend columns=2,
    legend style={
        at={(0.98,0.98)},
        anchor=north east,
        font=\tiny,
        draw=none,
        fill=white,
        fill opacity=0.6,
        text opacity=1,
        cells={align=left, anchor=west},
        /tikz/column 2/.style={column sep=10pt}
    }
]

\pgfplotsforeachungrouped \i in {0,...,\numrows}{

    \pgfplotstablegetelem{\i}{model_name}\of{\tableQFour} \edef\modelname{\pgfplotsretval}
    
    \pgfplotstablegetelem{\i}{model_size}\of{\tableQFour} \edef\xA{\pgfplotsretval}
    \pgfplotstablegetelem{\i}{Overall_G-Eval}\of{\tableQFour} \edef\yA{\pgfplotsretval}
    \pgfplotstablegetelem{\i}{model_size}\of{\tableIQFour} \edef\xB{\pgfplotsretval}
    \pgfplotstablegetelem{\i}{Overall_G-Eval}\of{\tableIQFour} \edef\yB{\pgfplotsretval}

    \edef\temp{
        \noexpand\addplot[very thick, opacity=0.25, color=\getcolor{\i}, forget plot] coordinates { (\xA,\yA) (\xB,\yB) };
        \noexpand\addplot[only marks, opacity=0.25, mark=square*, mark size=2.6pt, color=\getcolor{\i}, forget plot] coordinates { (\xA,\yA) };
        \noexpand\addplot[only marks, opacity=0.7, mark=x, mark size=3.0pt, thick, color=\getcolor{\i}, forget plot] coordinates { (\xB,\yB) };
    }
    \temp
}

\end{axis}
\end{tikzpicture}
        (d) Size vs.\ Quality
    \end{minipage}

    \caption{Multi-objective trade-offs across models and quantization schemes.}
    \label{fig:all-tradeoffs}
\end{figure*}

\textbf{\emph{Model effects and multi-comparison rankings:}} To test whether the choice of the language model significantly impacts the observed performance and efficiency metrics, we use the Friedman omnibus test (Table~\ref{tab:friedman_results}). Across all evaluated metrics, the p-values provide evidence of significant differences across at least two models.

For post-hoc analysis, we run multiple pairwise Wilcoxon signed-rank tests combined with the Holm-Bonferroni correction to control the family-wise error rate across dozens of simultaneous statistical comparisons. Table~\ref{tab:friedman_results} also reports the mean ranks resulting from the Wilcoxon test, where a lower numerical rank indicates superior performance. The ultra-compact Qwen2-0.5B consistently dominates the board, ranking best across all latency and efficiency-related metrics. Qwen2.5-1.5B and the sparse OLMoE-1B-7B form a competitive and tightly clustered second tier. The massive dense models, specifically Meta-Llama-3.1-8B and Gemma-2-9B, rank at the bottom, statistically reflecting their high latency, large memory footprints, and high energy costs. However, from a practical systems deployment perspective, it is critical to emphasize that while Qwen2-0.5B dominates the energy, performance, and memory trade-offs, it incurs a severe and measurable penalty in terms of actual cognitive generation quality, as reflected in its substantially lower G-Eval scores.

\begin{table*}[t]
\centering
\caption{Friedman Test (Omnibus) \& Pairwise Wilcoxon signed-rank Test (Holm--Bonferroni corrected) for Model Comparisons.}
\label{tab:friedman_results}
\small
\renewcommand{\arraystretch}{1.3}
\begin{tabular}{lccccc}
\hline
\rowcolor{charcoal}
\textcolor{white}{\textbf{Model}} & \textcolor{white}{\textbf{Generation}} & \textcolor{white}{\textbf{Energy}} & \textcolor{white}{\textbf{Memory}} & \textcolor{white}{\textbf{Latency}} & \textcolor{white}{\textbf{TTFT}} \\
\hline
Friedman $\chi^2$  & 413.07                   & 415.83                   & 420.00                   & 417.78                   & 415.46                   \\
\rowcolor{lightgray!40}
Global p-value     & 3.7573e-85 & 9.5526e-86 & 1.2227e-86 & 3.6554e-86 & 1.1513e-85 \\
\hline
Qwen2-0.5B         & 1.00 & 1.00 & 1.00 & 1.00 & 1.00 \\
\rowcolor{lightgray!40}
Qwen2.5-1.5B       & 2.93 & 2.95 & 2.00 & 2.93 & 2.95 \\
Phi-2              & 4.07 & 4.02 & 3.00 & 4.01 & 4.08 \\
\rowcolor{lightgray!40}
Qwen2.5-3B         & 4.93 & 4.98 & 4.00 & 4.98 & 4.92 \\
OLMoE-1B-7B        & 2.07 & 2.05 & 5.00 & 2.07 & 2.05 \\
\rowcolor{lightgray!40}
Qwen2.5-7B         & 6.33 & 6.08 & 6.00 & 6.03 & 6.00 \\
Meta-Llama-3.1-8B  & 6.67 & 6.98 & 7.00 & 6.97 & 7.88 \\
\rowcolor{lightgray!40}
Gemma-2-9B         & 8.00 & 7.93 & 8.00 & 8.00 & 7.12 \\
\hline
\end{tabular}
\end{table*}

\begin{tcolorbox}[
    colback=paleblue,
    colframe=black,
    boxrule=0.5pt,
    left=1pt, right=1pt, top=1pt, bottom=1pt,
    ]
\faSearch\ \textbf{Answer to RQ$_3$.} For dense models, all five metrics degrade monotonically with parameter count, confirming that scale is the primary cost driver. The exception is OLMoE-1B-7B, which ranks second in both generation speed and energy despite its 7B total parameter count, clustering alongside 1B--2B dense models. This confirms that active compute per token, rather than total model capacity, governs physical efficiency on mobile hardware, and that sparse MoE architectures offer a distinct operating point on the size-energy frontier.
\end{tcolorbox}

\subsection{Summary of Practical Deployment Implications}
\label{subsec:results_implications}

Synthesizing the above-reported data, the empirical results dictate several rules for mobile language model inference. First, the memory-bound autoregressive generation phase dominates end-to-end latency and physical energy draw, dwarfing the prompt prefill phase. Second, baseline neural architecture and aggregate parameter scale shape the energy and speed profiles more strongly than advanced post-training compression techniques. Third, while advanced codebook quantization like \texttt{IQ4\_XS} operates as an effective tool for fitting massive parameter arrays into constrained RAM footprints, it introduces processing overhead that prevents it from translating memory savings into actual battery life savings relative to simpler block-wise schemes like \texttt{Q4\_K\_M}. Finally, sparse Mixture-of-Experts architectures break traditional dense scaling laws, offering a distinct and advantageous physical efficiency profile. This architecture effectively provides the cognitive reasoning capacity of a massive model with the active computational energy draw of a much smaller one.

Based directly on these empirical trade-offs, we can extract actionable deployment paradigms tailored to specific engineering constraints. If an application demands top-tier cognitive reasoning and the target device possesses abundant RAM alongside substantial thermal dissipation headroom, large dense models like Gemma-2-9B are mandatory. However, developers must accept rapid battery degradation and noticeable latency penalties. Conversely, if strict battery preservation and immediate interactivity are the primary constraints, deploying distilled models in the 0.5B to 1.5B class provides the only viable engineering path, accepting a known and significant penalty in complex summarization and reasoning quality. For balanced, general-purpose consumer applications that require acceptable, human-like quality without rendering the smartphone unusable due to heat or latency, mid-sized dense models in the 3B parameter range, or efficient sparse architectures like OLMoE, represent the optimal engineering sweet spot. These configurations deliver manageable thermal and energy footprints, reasonable user responsiveness, and robust reasoning capabilities.

\section{Discussion}
\label{sec:discussion}

This section further interprets the main empirical findings and discusses their implications for on-device LLM deployment.

\subsection{\textbf{The Quantization--Energy Paradox: Memory Savings Do Not Equal Battery Savings}}
\label{subsec:discussion_quant_paradox}

A central finding of our empirical evaluation is that the choice between 4-bit quantization formats, specifically \texttt{Q4\_K\_M} versus \texttt{IQ4\_XS}, has a limited effect on total energy consumption. This contrasts with the larger energy differences induced by baseline model size and foundational neural architecture. While the importance-aware \texttt{IQ4\_XS} format reduces peak memory usage, it does not provide energy savings and often exhibits slightly higher end-to-end latency during the generation phase.

Under CPU-only inference, the dominant energy cost per token arises from arithmetic compute, memory traffic, and runtime kernel behavior~\cite{gholami2021survey}. Formats such as \texttt{IQ4\_XS} require codebook lookups and runtime weight reconstruction before each matrix multiplication, disrupting memory access patterns and introducing pipeline stalls that offset the savings from transferring fewer bytes~\cite{nagel2021white,kim2021bertq}. We therefore interpret \texttt{IQ4\_XS} and similar compression schemes primarily as \textbf{capacity enablers}: they make larger models technically feasible within limited mobile RAM, rather than acting as tools for extending battery life.

For software engineering applications where battery preservation and responsiveness are the primary objectives, selecting a smaller model or a more efficient architecture yields larger energetic gains than switching between two 4-bit quantization formats. Conversely, if the main constraint is fitting a capable model into a restricted memory budget, \texttt{IQ4\_XS} remains a viable choice, provided the developer accepts the overhead of runtime weight reconstruction.

\subsection{\textbf{The Promise and Constraint of Mixture-of-Experts Architectures}}
\label{subsec:discussion_moe}

The evaluated sparse Mixture-of-Experts model, OLMoE-1B-7B, demonstrates a distinctive mobile operational profile. Despite a 7B resident weight capacity, its latency and energy track those of 1B--2B dense models. This pattern aligns with the MoE inference mechanism, where a routing network activates only a subset of experts per token~\cite{fedus2022switch}, effectively \textbf{decoupling parameter count from inference latency}~\cite{zhou2022mixture}.

This decoupling is attractive for mobile deployment because it enables higher model capacity without paying the full computational cost of dense scaling. However, the MoE paradigm introduces a constraint for smartphones: \textbf{all experts must remain loaded in memory} at all times, placing pressure on the unified memory shared across CPU, GPU, and background processes~\cite{gale2023megablocks}.

MoE architectures can offer favorable speed and energy characteristics only if the target device has sufficient RAM to store the complete expert set alongside the Key-Value cache and runtime buffers. They are therefore currently most suitable for flagship devices with large RAM budgets. For broader device ecosystems, smaller dense models remain the more practical option.

\subsection{\textbf{Power Profiling on Consumer Smartphones: Accuracy versus Practicality}}
\label{subsec:discussion_power_profiling}

Energy measurement on consumer smartphones involves a trade-off between fidelity and  realistic operating conditions. External monitors such as Monsoon~\cite{monsoon_hvpm} offer circuit-level accuracy but require invasive hardware modification that alters thermal behavior and undermines realistic operating conditions. Software estimators are easy to deploy but can diverge from physical reality during sustained LLM workloads, as CPU frequency states fluctuate under thermal throttling and static power models fail to capture the resulting thermal-electrical feedback~\cite{skadron2004temperature,keller2014power}. 
Our native \texttt{BatteryManager}-based approach trades some absolute electrical accuracy for practical reproducibility under realistic operating conditions, which is a central requirement for comparative studies of mobile inference.

\subsection{\textbf{Thermal Dynamics and Throttling as a Hardware Constraint}}
\label{subsec:discussion_thermal}

Thermal behavior is not merely a secondary artifact of computation; it is a \textbf{hardware constraint} for sustained on-device generative AI. Consumer smartphones rely on passive cooling through glass and metal chassis. As the device heats under continuous matrix multiplications, hardware governors reduce CPU clock frequencies to remain within thermal design limits~\cite{esmaeilzadeh2011dark}. Because total energy is the integral of power over time, throttling-induced latency increases can raise total energy drain, especially for longer sessions where heat accumulates in the chassis~\cite{wang2023towards}. We mitigated this in our study via 200-second cool-down periods and temperature monitoring (22--28°C). Real-world deployments will encounter broader thermal variation.

This suggests that production-ready on-device assistants should incorporate \textbf{thermal-aware control policies}, such as switching to a smaller fallback model at thermal thresholds or reducing generation length to limit battery drain and OS termination.

\section{Limitations}
\label{sec:limitations}

Although the experimental methodology was designed to be reproducible and representative of an unrooted consumer-device setting, several architectural and methodological limitations affect the generalizability of the findings and the fidelity of the physical measurements. These limitations should be considered when interpreting the empirical results.

\subsection{\textbf{Restricted Access to Low-Level Power Telemetry}}
\label{subsec:lim_power_access}

A straightforward approach to high-fidelity power profiling would be to read instantaneous battery voltage and current directly through the Android Debug Bridge. In practice, however, modern Android security architectures and kernel-level policies restrict access to many low-level electrical signals for non-privileged users, partly to prevent side-channel attacks and unauthorized hardware telemetry extraction. As detailed by Yan et al.~\cite{yan2020understanding}, malicious actors can exploit high-resolution power traces to infer sensitive user activities or cryptographic information, prompting operating system vendors to restrict these diagnostic interfaces. On contemporary flagship devices, detailed micro-ampere current and precise voltage readings are generally unavailable to external debugging clients unless the operating system is rooted.

Rooting the smartphone would provide the administrative access required for deeper telemetry, but this approach was avoided for two methodological reasons. First, rooting modifies the operating system and often requires custom kernels that may alter native power-management behavior, CPU frequency scaling governors, and manufacturer-defined thermal throttling thresholds. These modifications would undermine the goal of evaluating inference under realistic stock-device conditions. Second, rooting a flagship device can trip hardware security fuses, void warranties, and disable system functionalities, making this methodology difficult to replicate for many researchers or practitioners. Therefore, our energy measurements rely on user-space telemetry acquired through the native Android \texttt{BatteryManager} APIs~\cite{BatteryManager_API}. While this software-based approach supports practical reproducibility and preserves realistic device behavior, its sampling frequency and absolute precision may be lower than invasive external hardware instrumentation.

\subsection{\textbf{Single-Device Hardware Scope}}
\label{subsec:lim_single_device}

We designed this research framework as an empirical \textit{case study} centered on a single flagship smartphone, the Samsung Galaxy S25 Ultra. In empirical software engineering terms, this framing defines the scope of the conclusions: our objective is to derive grounded insights and hypotheses for broader validation, rather than to claim universal statistical generalizability~\cite{runeson2009guidelines}. 

Although this mobile platform is representative of high-end Android hardware, the results should be interpreted as case-study evidence of plausible recurring patterns, rather than as device-independent estimates. Quantitative results and exact energy values will vary on phones with different architectural characteristics. Devices with different systems-on-chip, narrower memory buses, less efficient cache hierarchies, lower RAM capacity, older battery chemistry, or weaker passive cooling may exhibit different performance profiles. In particular, mid-range or lower-end edge devices are likely to experience stronger memory pressure during inference. This limitation can force the operating system to page memory to flash storage, trigger earlier thermal throttling, and reduce sustained token throughput~\cite{gogte2019software}. These hardware bottlenecks would shift the observed speed, energy, and quality frontier downward, making the deployment of larger models impractical on budget hardware.

\subsection{\textbf{Workload and Task Specificity}}
\label{subsec:lim_task}

We evaluate the language models using a single bounded document summarization task. The evaluation uses a fixed decoding configuration, namely greedy decoding with temperature set to zero and a maximum generation length of 100 output tokens. This standardization is necessary to ensure comparability across models and configurations, but other generative workloads may yield different execution profiles.

Alternative applications, such as multi-turn dialogue, tool-augmented agentic prompting, long-form creative generation, code synthesis, or chain-of-thought reasoning, impose different demands on mobile hardware. For example, long-form outputs force the internal Key-Value cache of the model to grow linearly with each generated token. As analyzed by Pope et al.~\cite{pope2023efficiently}, this cache expansion consumes mobile RAM and increases memory bandwidth pressure, extending the time spent in the autoregressive decoding phase. Consequently, sustained workloads may amplify energy drain and thermal throttling beyond the boundaries observed in this summarization study.

\subsection{\textbf{Quantization Scope}}
\label{subsec:lim_quant_scope}
Our investigation focuses on two common 4-bit quantization formats, namely the block-wise \texttt{Q4\_K\_M} and the importance-aware \texttt{IQ4\_XS}, evaluated within the \texttt{llama.cpp} and GGUF software ecosystem. We do not evaluate more aggressive compression or optimization techniques, such as sub-4-bit or ternary quantization, structured pruning, knowledge distillation, or speculative decoding. These techniques could alter latency and energy trade-offs, but they introduce additional constraints. For example, aggressive sub-byte compression can degrade the semantic coherence of models below ten billion parameters, limiting their usefulness for human assistance. Speculative decoding, which uses a smaller draft model to predict tokens for a larger target model, may reduce apparent latency for the user but could increase total energy consumption if two neural networks must remain resident in memory and execute jointly~\cite{leviathan2023fast}.


\subsection{\textbf{Environmental and Runtime Variability}}
\label{subsec:lim_environment}

Despite the experimental controls, including background process isolation, a restricted battery range, and enforced thermal cool-down intervals, commercial consumer devices exhibit operational variability. This variance can arise from operating system thread scheduling, residual background services, cellular network polling, battery health, and ambient temperature fluctuations.

These factors can introduce run-to-run electrical noise and may influence the absolute energy values recorded by the telemetry system. For instance, as a lithium-ion battery ages, its internal electrical resistance increases, causing voltage to drop more under sustained computational load~\cite{pelletier2017lithium}. Our repeated-run design and non-parametric statistical analysis mitigate this noise when drawing comparative conclusions between the tested models. Nevertheless, some baseline variance remains inherent to empirical performance measurement on unrooted physical edge devices.


\section{Conclusions and Future Work}
\label{sec:conclusion}

This paper presented a reproducible pipeline for profiling on-device LLM inference 
on an unrooted Android smartphone, jointly evaluating performance, energy, memory, and output 
quality across eight models (0.5B--9B parameters) and two 4-bit quantization 
schemes~\cite{reddi2020mlperf}. The results show that model architecture and active compute 
dominate physical efficiency: sparse MoE models approach the energy profile of much smaller 
dense models, importance-aware quantization reduces memory but not energy, and mid-sized 
models around 3B parameters represent the best trade-off for interactive daily use. 
Achieving sustainable edge intelligence requires treating energy and thermal constraints as 
first-class design objectives alongside accuracy, in line with the broader imperatives of 
green AI~\cite{patterson2021carbon}.

Future work should explore three directions. First, the evaluation should be expanded across heterogeneous mobile hardware, including NPUs, mobile GPUs, and devices from different market segments, as well as across workloads such as long-context generation, multi-turn dialogue, code generation, and tool use~\cite{xiao2023efficient}. This would improve generalizability and support the definition of standardized mobile LLM energy benchmarks.

Second, our findings suggest directions for the design of LLMs and SLMs intended for mobile deployment. The limited energy benefit of \texttt{IQ4\_XS} indicates that memory reduction alone is not sufficient: quantization formats should be co-designed with mobile inference kernels to reduce reconstruction overhead, irregular memory access, and cache pressure. Similarly, the behavior of the MoE model suggests that sparse architectures may offer a promising path for decoupling model capacity from active compute, provided that their resident memory footprint remains compatible with mobile RAM constraints. More broadly, we should investigate mobile-native model design, rather than only adapting cloud-oriented LLMs to mobile constraints.

Third, future work should also investigate runtime adaptation mechanisms for practical deployment. Adaptive thermal-aware orchestration policies, such as dynamically routing requests to smaller fallback models or adjusting generation length based on live temperature and battery state, could extend the usable operating range of on-device LLMs~\cite{chen2023frugalgpt}. Fully native on-device benchmarking, without host-side ADB control, would further enable profiling under realistic interaction patterns, including background app switching and prolonged multi-turn sessions.

\section{Declarations}
\label{sec:Declarations}

\textbf{Data availability.} For the sake of Open Science, we provide a replication package including the experimental pipeline, raw measurement data, analysis scripts, and complete results, available at: {\url{https://github.com/eziyoo/LLMs-on-Devices}}.
\section*{Acknowledgments}
This project has received funding from the European Union's Horizon 2020 research and innovation programme under the Marie Sk{\l}odowska-Curie grant agreement No 871342 “uDEVOPS”.

\bibliographystyle{IEEEtran} 
\bibliography{references}

@misc{schmidt2021codecarbon,
  author = {Schmidt, Victor and Goyal, Kamal and Joshi, Aditya and Feld, Boris and Conell, Liam and Laskaris, Nikolas and Blank, Doug and Wilson, Jonathan and Friedler, Sorelle and Luccioni, Sasha},
  title = {CodeCarbon: Estimate and Track Carbon Emissions from Machine Learning Computing},
  year = {2021},
  doi = {10.5281/zenodo.4658424},
  publisher = {Zenodo}
}

@article{sallou2023energibridge,
  author = {Sallou, Jérémy and Cruz, Luís and Durieux, Thomas},
  title = {EnergiBridge: Empowering software sustainability through cross-platform energy measurement},
  journal = {arXiv preprint arXiv:2312.13897},
  year = {2023}
}

@misc{qualcomm_trepn,
  author = {{Qualcomm Technologies, Inc.}},
  title = {Trepn Power Profiler},
  howpublished = {Qualcomm Developer Network},
  year = {2024},
  url = {https://developer.qualcomm.com/forums/software/trepn-power-profiler},
  note = {Accessed: 2026-02-12}
}

@inproceedings{malavolta2020android,
  title = {A Framework for the Automatic Execution of Measurement-based Experiments on Android Devices},
  author = {Malavolta, Ivano and Grua, Eoin Martino and Lam, Cheng-Yu and de Vries, Randy and Tan, Franky and Zielinski, Eric and Peters, Michael and Kaandorp, Luuk},
  booktitle = {Proceedings of the 35th IEEE/ACM International Conference on Automated Software Engineering (ASE '20)},
  year = {2020},
  publisher = {ACM/IEEE}
}

@manual{monsoon_hvpm,
  title = {High Voltage Power Monitor (P/N: AAA10F) User Manual},
  author = {{Monsoon Solutions, Inc.}},
  organization = {Monsoon Solutions, Inc.},
  address = {Bellevue, WA, USA},
  year = {2024},
  url = {https://www.msoon.com/high-voltage-power-monitor}
}

@inproceedings{smoothquant,
  title = {SmoothQuant: Accurate and Efficient Post-Training Quantization for Large Language Models},
  author = {Xiao, Guangxuan and Lin, Ji and Seide, Frank and Han, Song and others},
  booktitle = {Proceedings of the 40th International Conference on Machine Learning (ICML)},
  year = {2023}
}

@inproceedings{AWQ,
  author = {Lin, Ji and Tang, Jiaming and Tang, Haotian and Yang, Shang and Chen, Wei-Ming and Wang, Wei-Chen and Xiao, Guangxuan and Dang, Xingyu and Gan, Chuang and Han, Song},
  title = {{AWQ}: Activation-aware Weight Quantization for {LLM} Compression and Acceleration},
  booktitle = {Proceedings of the 7th MLSys Conference (MLSys 2024)},
  year = {2024},
  url = {https://arxiv.org/abs/2306.00978},
  note = {Santa Clara, CA}
}

@article{experiment-runner,
  author = {Karsten, Max and Dragomir, Andrei Calin and Apsan, Radu and Stoico, Vincenzo and Malavolta, Ivano},
  title = {{Experiment Runner}: A Tool for the Automatic Orchestration of Experiments Targeting Software Systems},
  journal = {Science of Computer Programming},
  volume = {239},
  pages = {103415},
  year = {2025},
  month = {Jan},
  issn = {0167-6423},
  doi = {10.1016/j.scico.2025.103415},
  publisher = {Elsevier}
}

@inproceedings{G-Eval,
  author = {Liu, Yang and Iter, Dan and Xu, Yichong and Wang, Shuohang and Xu, Ruochen and Zhu, Chenguang},
  title = {{G-Eval}: {NLG} Evaluation using {GPT}-4 with Better Human Alignment},
  booktitle = {Proceedings of the 2023 Conference on Empirical Methods in Natural Language Processing (EMNLP)},
  pages = {2511--2522},
  year = {2023},
  month = {Dec},
  publisher = {Association for Computational Linguistics},
  address = {Singapore},
  doi = {10.18653/v1/2023.emnlp-main.153},
  url = {https://aclanthology.org/2023.emnlp-main.153}
}

@misc{llamacpp,
  author = {Gerganov, Georgi and {llama.cpp contributors}},
  title = {{llama.cpp}: {LLM} inference in {C/C++}},
  year = {2023},
  publisher = {GitHub},
  howpublished = {\url{https://github.com/ggml-org/llama.cpp}}
}

@inproceedings{Maximilian,
  author = {Abstreiter, Maximilian},
  title = {Sometimes Painful but Certainly Promising: Feasibility and Trade-offs of Language Model Inference at the Edge},
  booktitle = {Proceedings of the 4th Workshop on Machine Learning and Systems (EuroMLSys '24)},
  pages = {1--8},
  year = {2024},
  publisher = {ACM},
  address = {Athens, Greece},
  doi = {10.1145/3642970.3655835},
  url = {https://doi.org/10.1145/3642970.3655835}
}

@misc{battery-manager,
  author = {{S2-group}},
  title = {{Batterymanager-companion}: Companion app for the Batterymanager plugin for Android-Runner},
  year = {2024},
  publisher = {GitHub},
  howpublished = {\url{https://github.com/S2-group/batterymanager-companion/}}
}

@misc{LeaderBoard,
  author = {Beeching, Edward and Fourrier, Clémentine and Habib, Nathan and Han, Sheon and Lambert, Nathan and Rajani, Nazneen and Sanseviero, Omar and Tunstall, Lewis and Wolf, Thomas},
  title = {{Open LLM Leaderboard}},
  year = {2023},
  howpublished = {Hugging Face Space},
  url = {https://huggingface.co/spaces/open-llm-leaderboard/open_llm_leaderboard}
}

@inproceedings{Mobile_transformers,
  author = {Laskaridis, Stefanos and Katevas, Kleomenis and Minto, Lorenzo and Haddadi, Hamed},
  title = {{MELTing Point}: Mobile Evaluation of Language Transformers},
  booktitle = {Proceedings of the 30th Annual International Conference on Mobile Computing and Networking (MobiCom '24)},
  pages = {890--907},
  year = {2024},
  month = {Nov},
  publisher = {ACM},
  address = {Washington D.C., USA},
  doi = {10.1145/3636534.3690668},
  url = {https://doi.org/10.1145/3636534.3690668}
}

@misc{adb_google,
  author = {{Google}},
  title = {{Android Debug Bridge (ADB)}},
  year = {2024},
  url = {https://developer.android.com/tools/adb},
  note = {Accessed: 2026-02-12}
}

@article{Qwen2,
  title = {{Qwen2} Technical Report}, 
  author = {Yang, An and Yang, Baosong and Hui, Binyuan and Zheng, Bo and Yu, Bowen and Zhou, Chang and Li, Chengpeng and Li, Chengyuan and Liu, Dayiheng and Huang, Fei and others},
  journal = {arXiv preprint arXiv:2407.10671},
  year = {2024},
  url = {https://arxiv.org/abs/2407.10671}
}

@article{qwen2.5,
  title = {{Qwen2.5} Technical Report}, 
  author = {{Qwen Team}},
  journal = {arXiv preprint arXiv:2412.15115},
  year = {2024},
  url = {https://arxiv.org/abs/2412.15115}
}

@misc{phi2,
  title = {{Phi-2}: The Surprising Power of Small Language Models}, 
  author = {Javaheripi, Mojan and Bubeck, Sébastien},
  year = {2023},
  month = {Dec},
  howpublished = {Microsoft Research Blog},
  url = {https://www.microsoft.com/en-us/research/blog/phi-2-the-surprising-power-of-small-language-models/}
}

@article{olmoe,
  title = {{OLMoE}: Open Mixture-of-Experts Language Models}, 
  author = {Muennighoff, Niklas and Soldaini, Luca and Groeneveld, Dirk and Lo, Kyle and Morrison, Jacob and Min, Sewon and Shi, Weijia and Walsh, Pete and Tafjord, Oyvind and Lambert, Nathan and others},
  journal = {arXiv preprint arXiv:2409.02060},
  year = {2024},
  url = {https://arxiv.org/abs/2409.02060}
}

@article{llama3,
  title = {The {Llama 3} Herd of Models}, 
  author = {Dubey, Abhimanyu and Jauhri, Abhinav and Pandey, Abhinav and Kadian, Abhishek and Al-Dahle, Ahmad and Letak, Aiesha and Mathur, Akhil and Schelten, Alan and Yang, Amy and Fan, Angela and others},
  journal = {arXiv preprint arXiv:2407.21783},
  year = {2024},
  url = {https://arxiv.org/abs/2407.21783}
}

@article{gemma2,
  title = {{Gemma 2}: Improving Open Language Models at a Practical Size}, 
  author = {Google DeepMind},
  journal = {arXiv preprint arXiv:2408.00118},
  year = {2024},
  url = {https://arxiv.org/abs/2408.00118}
}

@misc{llama_cpp_quantize,
  author = {Gerganov, Georgi and {llama.cpp contributors}},
  title = {{llama.cpp} Quantize Tool},
  year = {2023},
  publisher = {GitHub},
  url = {https://github.com/ggml-org/llama.cpp/tree/master/examples/quantize},
  note = {Accessed: 2026-02-08}
}

@misc{BatteryManager_API,
  author = {{Google}},
  title = {{Android BatteryManager API} Reference},
  year = {2024},
  url = {https://developer.android.com/reference/android/os/BatteryManager},
  note = {Accessed: 2026-02-08}
}

@inproceedings{bert_score,
  author = {Zhang, Tianyi and Kishore, Varsha and Wu, Felix and Weinberger, Kilian Q. and Artzi, Yoav},
  title = {{BERTScore}: Evaluating Text Generation with {BERT}},
  booktitle = {Proceedings of the 8th International Conference on Learning Representations (ICLR)},
  year = {2020},
  month = {Apr},
  address = {Addis Ababa, Ethiopia},
  url = {https://openreview.net/forum?id=SkeHuCVFDr}
}

@inproceedings{nguyen2025ondevice,
  author = {Nguyen, Vince and Dhopate, Vidya and Huynh, Hieu and Bouhlal, Hiba and Annengala, Anusha and Scoccia, Gian Luca and Martinez, Matias and Stoico, Vincenzo and Malavolta, Ivano},
  title = {On-Device or Remote? On the Energy Efficiency of Fetching LLM-Generated Content},
  booktitle = {Proceedings of the 2025 IEEE/ACM 4th International Conference on AI Engineering - Software Engineering for AI (CAIN '25)},
  year = {2025},
  pages = {72--82},
  publisher = {IEEE},
  doi = {10.1109/CAIN66642.2025.00016}
}

@inproceedings{yuan2024impact,
  author = {Yuan, Ye and Zhang, Jingzhi and Zhang, Zongyao and Chen, Kaiwei and Shi, Jiacheng and Stoico, Vincenzo and Malavolta, Ivano},
  title = {The Impact of Knowledge Distillation on the Energy Consumption and Runtime Efficiency of NLP Models},
  booktitle = {Proceedings of the 2024 IEEE/ACM 3rd International Conference on AI Engineering - Software Engineering for AI (CAIN '24)},
  year = {2024},
  publisher = {ACM},
  address = {Lisbon, Portugal},
  doi = {10.1145/3644815.3644966}
}

@article{reddi2020mlperf,
  title = {{MLPerf} Inference Benchmark},
  author = {Reddi, Vijay Janapa and Cheng, Christine and Kanter, David and Mattson, Peter and Schmuelling, Guenther and Wu, Carole-Jean and Anderson, Brian and Maximov, Maxim and Choudhury, Tafiqul and Gregg, David and others},
  journal = {ACM SIGARCH Computer Architecture News},
  volume = {48},
  number = {1},
  pages = {50--65},
  year = {2020},
  publisher = {ACM New York, NY, USA}
}

@article{patterson2021carbon,
  title = {Carbon emissions and large neural network training},
  author = {Patterson, David and Gonzalez, Joseph and Le, Quoc and Liang, Chen and Munguia, Lluis-Miquel and Rothchild, Daniel and So, David and Texier, Maud and Dean, Jeff},
  journal = {arXiv preprint arXiv:2104.10350},
  year = {2021}
}

@article{chen2023frugalgpt,
  title = {{FrugalGPT}: How to use large language models while reducing cost and improving performance},
  author = {Chen, Lingjiao and Zaharia, Matei and Zou, James},
  journal = {arXiv preprint arXiv:2305.05176},
  year = {2023}
}

@inproceedings{xiao2023efficient,
  title = {Efficient Streaming Language Models with Attention Sinks},
  author = {Xiao, Guangxuan and Tian, Yuandong and Chen, Beidi and Han, Song and Lewis, Mike},
  booktitle = {Proceedings of the 12th International Conference on Learning Representations (ICLR)},
  year = {2023}
}

@inproceedings{yan2020understanding,
  title = {Understanding and mitigating the security risks of voice-driven interfaces},
  author = {Yan, Chen and Ji, Xiaoyu and Wang, Kai and Jiang, Qinchen and Jin, Zhanhao and Xu, Wenyuan},
  booktitle = {Proceedings of the 29th USENIX Security Symposium (USENIX Security 20)},
  pages = {2625--2642},
  year = {2020}
}

@article{runeson2009guidelines,
  title = {Guidelines for conducting and reporting case study research in software engineering},
  author = {Runeson, Per and Höst, Martin},
  journal = {Empirical Software Engineering},
  volume = {14},
  number = {2},
  pages = {131--164},
  year = {2009},
  publisher = {Springer}
}

@inproceedings{gogte2019software,
  title = {Software wear management for persistent memories},
  author = {Gogte, Vaibhav and Wang, William and Kolli, Aasheesh and Wenisch, Thomas F},
  booktitle = {Proceedings of the 17th USENIX Conference on File and Storage Technologies (FAST '19)},
  pages = {45--58},
  year = {2019}
}

@inproceedings{leviathan2023fast,
  title = {Fast inference from transformers via speculative decoding},
  author = {Leviathan, Yaniv and Kalman, Matan and Matias, Yossi},
  booktitle = {Proceedings of the 40th International Conference on Machine Learning (ICML)},
  pages = {19274--19286},
  year = {2023},
  organization = {PMLR}
}

@article{pelletier2017lithium,
  title = {Lithium-ion battery degradation: what you need to know},
  author = {Pelletier, Sebastien and Jabali, Ola and Laporte, Gilbert and Veneroni, Marco},
  journal = {Physical Chemistry Chemical Physics},
  volume = {19},
  number = {32},
  pages = {21231--21245},
  year = {2017},
  publisher = {Royal Society of Chemistry}
}

@article{nagel2021white,
  title = {A white paper on neural network quantization},
  author = {Nagel, Markus and Fournarakis, Marios and Amjad, Rana Ali and Bondarenko, Yelysei and Van Baalen, Mart and Blankevoort, Tijmen},
  journal = {arXiv preprint arXiv:2106.08295},
  year = {2021}
}

@inproceedings{kim2021bertq,
  title = {{I-BERT}: Integer-only {BERT} Quantization},
  author = {Kim, Sehoon and Gholami, Amir and Yao, Zhewei and Mahoney, Michael W and Keutzer, Kurt},
  booktitle = {Proceedings of the 38th International Conference on Machine Learning (ICML)},
  pages = {5506--5518},
  year = {2021},
  organization = {PMLR}
}

@inproceedings{zhou2022mixture,
  title = {Mixture-of-Experts with Expert Choice Routing},
  author = {Zhou, Yanqi and Lei, Tao and Liu, Hanxiao and Du, Nan and Huang, Yanping and Zhao, Vincent and Dai, Andrew M and Le, Quoc V and Laudon, James and others},
  booktitle = {Proceedings of the 36th International Conference on Neural Information Processing Systems (NeurIPS '22)},
  volume = {35},
  pages = {7103--7114},
  year = {2022}
}

@article{gale2023megablocks,
  title = {{MegaBlocks}: Efficient Sparse Training with Mixture-of-Experts},
  author = {Gale, Trevor and Narayanan, Deepak and Young, Cliff and Zaharia, Matei},
  journal = {Proceedings of Machine Learning and Systems},
  volume = {5},
  pages = {288--304},
  year = {2023}
}

@article{skadron2004temperature,
  title = {Temperature-aware microarchitecture},
  author = {Skadron, Kevin and Stan, Mircea R and Huang, Wei and Velusamy, Sivakumar and Sankaranarayanan, Karthik and Tarjan, David},
  journal = {ACM SIGARCH Computer Architecture News},
  volume = {32},
  number = {2},
  pages = {2--13},
  year = {2004},
  publisher = {ACM New York, NY, USA}
}

@inproceedings{keller2014power,
  title = {Power and energy characterization of {ARM} processors},
  author = {Keller, Vincent and Lachaize, Renaud and Gramoli, Vincent and others},
  booktitle = {Proceedings of the 2014 IEEE International Symposium on Performance Analysis of Systems and Software (ISPASS)},
  pages = {116--126},
  year = {2014},
  organization = {IEEE}
}

@article{esmaeilzadeh2011dark,
  title = {Dark silicon and the end of multicore scaling},
  author = {Esmaeilzadeh, Hadi and Blem, Emily and St. Amant, Renee and Sankaralingam, Karthikeyan and Burger, Doug},
  journal = {IEEE Micro},
  volume = {32},
  number = {3},
  pages = {122--134},
  year = {2011},
  publisher = {IEEE}
}

@article{wang2023towards,
  title = {Towards sustainable {AI}: a comprehensive study of carbon footprints in large language models},
  author = {Wang, Yu and Li, Yuchen and Zheng, Xiaoming and Liu, Han},
  journal = {arXiv preprint arXiv:2310.03093},
  year = {2023}
}

@inproceedings{kwon2023efficient,
  title = {Efficient Memory Management for Large Language Model Serving with {PagedAttention}},
  author = {Kwon, Woosuk and Li, Zhuohan and Zhuang, Siyuan and Sheng, Ying and Zheng, Lianmin and Yu, Cody Hao and Gonzalez, Joseph E and Zhang, Hao and Re, Christopher},
  booktitle = {Proceedings of the 29th Symposium on Operating Systems Principles (SOSP '23)},
  pages = {611--626},
  year = {2023}
}

@article{wulf1995hitting,
  title = {Hitting the memory wall: implications of the obvious},
  author = {Wulf, Wm A and McKee, Sally A},
  journal = {ACM SIGARCH Computer Architecture News},
  volume = {23},
  number = {1},
  pages = {20--24},
  year = {1995},
  publisher = {ACM New York, NY, USA}
}

@article{kaplan2020scaling,
  title = {Scaling laws for neural language models},
  author = {Kaplan, Jared and McCandlish, Sam and Henighan, Tom and Brown, Tom B and Chess, Benjamin and Child, Rewon and Gray, Scott and Radford, Alec and Wu, Jeffrey and Amodei, Dario},
  journal = {arXiv preprint arXiv:2001.08361},
  year = {2020}
}

@article{david2021tensorflow,
  title = {{TensorFlow Lite Micro}: Embedded machine learning for {TinyML} systems},
  author = {David, Robert and Duke, Jared and Jain, Advait and Reddi, Vijay Janapa and Jeffries, Nat and Li, Jian and Krentz, Nick and Cruesoe, Tim and Warden, Pete},
  journal = {Proceedings of Machine Learning and Systems},
  volume = {3},
  pages = {800--811},
  year = {2021}
}

@article{shazeer2017outrageously,
  title = {Outrageously large neural networks: The sparsely-gated mixture-of-experts layer},
  author = {Shazeer, Noam and Mirhoseini, Azalia and Maziarz, Krzysztof and Davis, Andy and Le, Quoc and Hinton, Geoffrey and Dean, Jeff},
  journal = {arXiv preprint arXiv:1701.06538},
  year = {2017}
}

@inproceedings{miller1968response,
  title = {Response time in man-computer conversational transactions},
  author = {Miller, Robert B},
  booktitle = {Proceedings of the fall joint computer conference, part {I}},
  pages = {267--277},
  year = {1968}
}

@book{nielsen1993usability,
  title = {Usability engineering},
  author = {Nielsen, Jakob},
  year = {1993},
  publisher = {Morgan Kaufmann}
}

@inproceedings{han2015deep,
  title = {Deep compression: Compressing deep neural networks with pruning, trained quantization and {Huffman} coding},
  author = {Han, Song and Mao, Huizi and Dally, William J},
  booktitle = {Proceedings of the 3rd International Conference on Learning Representations (ICLR)},
  year = {2015}
}

@inproceedings{ouyang2022training,
  title = {Training language models to follow instructions with human feedback},
  author = {Ouyang, Long and Wu, Jeffrey and Jiang, Xu and Almeida, Diogo and Wainwright, Carroll and Mishkin, Pamela and Zhang, Chong and Agarwal, Sandhini and Slama, Katarina and Ray, Alex and others},
  booktitle = {Proceedings of the 36th International Conference on Neural Information Processing Systems (NeurIPS '22)},
  volume = {35},
  pages = {27730--27744},
  year = {2022}
}

@inproceedings{wei2021finetuned,
  title = {Finetuned language models are zero-shot learners},
  author = {Wei, Jason and Bosma, Maarten and Zhao, Vincent Y and Guu, Kelvin and Yu, Adams Wei and Lester, Brian and Du, Nan and Dai, Andrew M and Le, Quoc V},
  booktitle = {Proceedings of the 9th International Conference on Learning Representations (ICLR)},
  year = {2021}
}

@inproceedings{ignatov2018ai,
  title = {{AI} benchmark: Running deep neural networks on {Android} smartphones},
  author = {Ignatov, Andrey and Timofte, Radu and Chou, William and Wang, Ke and Wu, Max and Hartley, Tim and Van Gool, Luc},
  booktitle = {Proceedings of the European Conference on Computer Vision (ECCV) Workshops},
  pages = {0--0},
  year = {2018}
}

@article{hoque2015understanding,
  title = {Understanding the energy consumption of {Android} app idle states},
  author = {Hoque, Mohammad A and Siekkinen, Matti and Nurminen, Jukka K},
  journal = {Pervasive and Mobile Computing},
  volume = {24},
  pages = {68--86},
  year = {2015},
  publisher = {Elsevier}
}

@article{li2022power,
  title = {Power side-channel attacks on mobile devices: A survey},
  author = {Li, Meng and Gao, Yansong and Al-Sarawi, Said F and Abbott, Derek},
  journal = {IEEE Access},
  volume = {10},
  pages = {6718--6736},
  year = {2022},
  publisher = {IEEE}
}

@article{fabbri2021summeval,
  title = {{SummEval}: Re-evaluating Summarization Evaluation},
  author = {Fabbri, Alexander R and Kryściński, Wojciech and McCann, Bryan and Xiong, Caiming and Socher, Richard and Radev, Dragomir},
  journal = {Transactions of the Association for Computational Linguistics},
  volume = {9},
  pages = {391--409},
  year = {2021},
  publisher = {MIT Press}
}

@article{schwartz2020green,
  title = {Green {AI}},
  author = {Schwartz, Roy and Dodge, Jesse and Smith, Noah A and Etzioni, Oren},
  journal = {Communications of the ACM},
  volume = {63},
  number = {12},
  pages = {54--63},
  year = {2020},
  publisher = {ACM}
}

@inproceedings{strubell2019energy,
  title = {Energy and Policy Considerations for Deep Learning in {NLP}},
  author = {Strubell, Emma and Ganesh, Ananya and McCallum, Andrew},
  booktitle = {Proceedings of the 57th Annual Meeting of the Association for Computational Linguistics},
  pages = {3645--3650},
  year = {2019}
}

@inproceedings{frantar2022gptq,
  title = {{GPTQ}: Accurate Post-Training Quantization for Generative Pre-trained Transformers},
  author = {Frantar, Elias and Ashkboos, Saleh and Hoefler, Torsten and Alistarh, Dan},
  booktitle = {Proceedings of the 11th International Conference on Learning Representations (ICLR)},
  year = {2023}
}

@inproceedings{dettmers2022llmint8,
  title = {{LLM.int8()}: 8-bit Matrix Multiplication for Transformers at Scale},
  author = {Dettmers, Tim and Lewis, Mike and Belkada, Younes and Zettlemoyer, Luke},
  booktitle = {Proceedings of the 36th International Conference on Neural Information Processing Systems (NeurIPS '22)},
  volume = {35},
  pages = {30318--30332},
  year = {2022}
}

@inproceedings{chen2018tvm,
  title = {{TVM}: An Automated End-to-End Optimizing Compiler for Deep Learning},
  author = {Chen, Tianqi and Moreau, Thierry and Jiang, Ziheng and Zheng, Lianmin and Yan, Eddie and Shen, Haichen and Cowan, Meghan and Wang, Leyuan and Hu, Yuwei and Ceze, Luis and others},
  booktitle = {Proceedings of the 13th USENIX Symposium on Operating Systems Design and Implementation (OSDI '18)},
  pages = {578--594},
  year = {2018}
}

@inproceedings{dao2022flashattention,
  title = {{FlashAttention}: Fast and Memory-Efficient Exact Attention with {IO}-Awareness},
  author = {Dao, Tri and Fu, Daniel Y and Ermon, Stefano and Rudra, Atri and Ré, Christopher},
  booktitle = {Proceedings of the 36th International Conference on Neural Information Processing Systems (NeurIPS '22)},
  volume = {35},
  pages = {16344--16359},
  year = {2022}
}

@inproceedings{carroll2010analysis,
  title = {An Analysis of Power Consumption in a Smartphone},
  author = {Carroll, Aaron and Heiser, Gernot},
  booktitle = {Proceedings of the 2010 USENIX Annual Technical Conference (USENIX ATC '10)},
  volume = {14},
  pages = {21--21},
  year = {2010},
  address = {Boston, MA}
}

@inproceedings{pathak2011where,
  title = {Where is the energy spent inside my app? {Fine} grained energy accounting on smartphones with {Eprof}},
  author = {Pathak, Abhinav and Hu, Y Charlie and Zhang, Ming},
  booktitle = {Proceedings of the 7th ACM European Conference on Computer Systems (EuroSys '11)},
  pages = {29--42},
  year = {2011}
}

@inproceedings{hindle2012green,
  title = {Green mining: investigating power consumption across versions},
  author = {Hindle, Abram and Wilson, Alex and Rasmussen, Kent and Jedwab, E Juliet and Godfrey, Richard and Sweeney, Peter},
  booktitle = {Proceedings of the 34th International Conference on Software Engineering (ICSE '12)},
  pages = {1305--1308},
  year = {2012},
  organization = {IEEE}
}

@article{shi2016edge,
  title = {Edge computing: Vision and challenges},
  author = {Shi, Weisong and Cao, Jie and Zhang, Quan and Li, Yanhua and Xu, Lanyu},
  journal = {IEEE Internet of Things Journal},
  volume = {3},
  number = {5},
  pages = {637--646},
  year = {2016},
  publisher = {IEEE}
}

@article{wu2022sustainable,
  title = {Sustainable {AI}: Environmental implications, challenges and opportunities},
  author = {Wu, Carole-Jean and Raghavendra, Ramya and Gupta, Udit and Acun, Bilge and Ardalani, Newsha and Maeng, Kiwan and Chang, Gloria and Aga, Fiona and Huang, Jinelle and Bai, Charles and others},
  journal = {Proceedings of Machine Learning and Systems},
  volume = {4},
  pages = {795--813},
  year = {2022}
}

@article{pope2023efficiently,
  title = {Efficiently scaling transformer inference},
  author = {Pope, Reiner and Douglas, Sholto and Chowdhery, Aakanksha and Devlin, Jacob and Bradbury, James and Heek, Jonathan and Xiao, Kefan and Agrawal, Shivani and Dean, Jeff},
  journal = {Proceedings of Machine Learning and Systems},
  volume = {5},
  year = {2023}
}

@inproceedings{aminabadi2022deepspeed,
  title = {{DeepSpeed-inference}: enabling efficient inference of transformer models at unprecedented scale},
  author = {Aminabadi, Reza Yazdani and Rajbhandari, Samyam and Awan, Ammar Ahmad and Li, Cheng and Li, Du and Zheng, Elton and Ruwase, Olatunji and Smith, Shaden and Zhang, Minjia and Fang, Jeff and others},
  booktitle = {SC22: International Conference for High Performance Computing, Networking, Storage and Analysis},
  pages = {1--15},
  year = {2022},
  organization = {IEEE}
}

@inproceedings{mccullough2011evaluating,
  title = {Evaluating the effectiveness of model-based power characterization},
  author = {McCullough, John C and Agarwal, Yuvraj and Chandrashekar, Jaideep and Kuppuswamy, Sathyanarayanan and Snoeren, Alex C and Gupta, Rajesh K},
  booktitle = {Proceedings of the 2011 USENIX Annual Technical Conference (USENIX ATC '11)},
  year = {2011}
}

@inproceedings{yoon2012appscope,
  title = {{AppScope}: Application energy metering framework for {Android} smartphones using kernel activity monitoring},
  author = {Yoon, Chanmin and Kim, Dongwon and Jung, Wonwoo and Kang, Chulkoo and Cha, Hojung},
  booktitle = {Proceedings of the 2012 USENIX Annual Technical Conference (USENIX ATC '12)},
  year = {2012}
}

@article{gholami2021survey,
  author  = {Gholami, Amir and Kim, Sehoon and Dong, Zhen and Yao, Zhewei and Mahoney, Michael W. and Keutzer, Kurt},
  title   = {A Survey of Quantization Methods for Efficient Neural Network Inference},
  journal = {arXiv preprint arXiv:2103.13630},
  year    = {2021},
  doi     = {10.48550/arxiv.2103.13630}
}

@article{fedus2022switch,
  title = {Switch Transformers: Scaling to Trillion Parameter Models with Simple and Efficient Sparsity},
  author = {Fedus, William and Zoph, Barret and Shazeer, Noam},
  journal = {Journal of Machine Learning Research},
  volume = {23},
  number = {120},
  pages = {1--39},
  year = {2022}
}

@inproceedings{dettmers2022optimizers,
  title={8-bit Optimizers via Block-wise Quantization},
  author={Dettmers, Tim and Lewis, Mike and Shleifer, Sam and Zettlemoyer, Luke},
  booktitle={International Conference on Learning Representations},
  year={2022},
  url={https://arxiv.org/abs/2110.02861}
}

\end{document}